\documentclass[letterpaper,12pt]{article}
\pdfoutput=1
\usepackage{color}
\usepackage[normalem]{ulem}
\usepackage{amsmath, amssymb, slashed}
\usepackage{bbm}
\usepackage{grffile,graphicx}
\usepackage{url}
\usepackage[dvipsnames]{xcolor}
\usepackage[utf8]{inputenc}
\usepackage[version=4]{mhchem}
\usepackage{mathtools}

\usepackage{tabularx} 
\usepackage[margin=1in,letterpaper]{geometry} 
\usepackage{cite} 
\usepackage[final,backref=page]{hyperref} 
\hypersetup{
	colorlinks=true,       
	linkcolor=blue,        
	citecolor=blue,        
	filecolor=magenta,     
	urlcolor=blue         
}
\usepackage{comment}
\usepackage{caption}
\usepackage{subcaption}
\usepackage{slashed}
\usepackage{bbm}
\usepackage{xcolor}
\usepackage{graphics}
\usepackage{authblk}
\usepackage{tikz}
\usepackage{afterpage}
    
\newcommand{\be}{\begin{equation}}
\newcommand{\ee}{\end{equation}}
\newcommand{\Tend}{T_\text{end}}
\newcommand{\Teq}{T_\text{eq}}

\newcommand{\Tc}{T_\text{c}}
\newcommand{\Rend}{R_\text{end}}

\newcommand{\Rc}{R_\text{c}}

\newcommand{\gs}{g_\star}
\newcommand{\gss}{g_{\star S}}
\newcommand{\rp}{\rho_\phi}
\newcommand{\rR}{\rho_R}

\newcommand{\Gphi}{\Gamma_\phi}

\newcommand{\ma}{m_a}
\newcommand{\Tqcd}{T_\text{QCD}}
\newcommand{\Tbbn}{T_\text{BBN}}

\newcommand{\Td}{T_\text{d}}
\newcommand{\Tdnsc}{T_\text{d,nsc}}

\newcommand{\DNeff}{\Delta N_\text{eff}}
\newcommand{\lfs}{\lambda_\text{fs}}

\begin{document}

\begin{center}
\vspace*{0.5cm}
{\Large{\textbf{Revisiting signatures of thermal axions\\in nonstandard cosmologies}}}
\date{}

\vspace*{1.2cm}

{\large
Paola Arias$^{a, b}$,\,%
Nicol\'as Bernal$^{c}$,\,%
Jacek K. Osiński$^{d}$,\\%
Leszek Roszkowski$^{d,e}$\,%
and Moira Venegas$^{a, f}$
}\\[3mm]
{\it{
$^{a}$ Departamento de Física, Universidad de Santiago de Chile\\
Casilla 307, Santiago, Chile\\
$^{b}$ Universidad San Sebastián, Santiago, Chile\\
$^{c}$ New York University Abu Dhabi\\
PO Box 129188, Saadiyat Island, Abu Dhabi, United Arab Emirates\\
$^{d}$ AstroCeNT, Nicolaus Copernicus Astronomical Center Polish Academy of Sciences\\
ul. Rektorska 4, 00-614 Warsaw, Poland\\
$^{e}$ National Centre for Nuclear Research\\
ul. Pasteura 7, 02-093 Warsaw, Poland\\
$^{f}$ Department of Physics and Astronomy, Rice University\\
Houston, Texas 77005, USA
}}
\end{center}

\begin{abstract}
    We revisit the formation of a thermal population of hadronic axions in nonstandard cosmologies, in light of the recent developments in obtaining continuous and smooth interaction rates for both the gluon and photon couplings. For certain cosmological histories, such as low-temperature reheating (LTR) and kination-like scenarios, the thermalization of the axion can be severely delayed to higher masses. In the case that thermal equilibrium is achieved, we improve the constraints on LTR for axion masses around the eV scale with respect to previous works and we constrain for the first time early matter-dominated cosmologies. We also point out the possibility of having the coexistence of cold and warm dark matter populations of axions in kination-like scenarios in the eV mass range.
\end{abstract}

\newpage
\tableofcontents
\newpage

\section{Introduction}
Axions are indisputably excellent cold dark matter (DM) candidates. They can be produced by the so-called misalignment mechanism, see, ({\it e.g.}, Refs.~\cite{Adams:2022pbo, DiLuzio:2020wdo, Marsh:2015xka}) or by the decay of topological defects such as cosmic strings and domain walls~\cite{Davis:1986xc, Hagmann:2000ja, Hiramatsu:2010yu, Kawasaki:2014sqa}, in a wide range of masses in the sub-eV ballpark. In the standard cosmological scenario (SC), axions can make up the DM density in the range of $\sim \mu$eV. However, this can be extended to smaller masses in scenarios such as kinetic misalignment~\cite{Co:2019jts, Chang:2019tvx, Barman:2021rdr}, or in nonstandard cosmologies (NSC) featuring periods with an injection of entropy~\cite{Allahverdi:2020bys}, typically due to the decay of a heavy long-lived particle~\cite{Steinhardt:1983ia, Lazarides:1990xp, Kawasaki:1995vt, Giudice:2000ex, Grin:2007yg, Visinelli:2009kt, Nelson:2018via, Visinelli:2018wza, Ramberg:2019dgi, Blinov:2019jqc, Carenza:2021ebx, Venegas:2021wwm, Arias:2021rer, Arias:2022qjt}, or by Hawking evaporation of primordial black holes~\cite{Bernal:2021yyb, Bernal:2021bbv, Choi:2022btl, Mazde:2022sdx}. However, if the misalignment mechanism occurred during a non-standard period such as kination domination (KD) (or in a more general scenario with an equation of state parameter $\omega > 1/3$), the observed DM relic abundance can be reproduced with masses near or around the eV scale~\cite{Visinelli:2009kt, Arias:2021rer}.

Furthermore, a thermal population of axions can also be produced from scatterings and/or decays of particles from the Standard Model (SM) to which they couple. In particular, a hot thermal population of axions forms because of the existence of the model-independent axion-gluon coupling. The effect of such a population depends on the mass of the axion. Axions with masses smaller than $\sim 0.1$~eV decouple from the SM thermal bath at temperatures above the QCD phase transition (QCDPT) and therefore experience an important entropy dilution due to the change in the relativistic degrees of freedom, or they may never reach thermal equilibrium at all, making them difficult to detect. Axions with higher masses that later decouple are more likely to have abundances that can be detected in our observations. In addition to the axion-gluon coupling, there is a coupling to photons; even though it is model-dependent, it is widely exploited to search for axions and axion-like particles. 

The hot axion population and its impact on observable cosmology attracted attention long ago. One of the first approaches to study their phenomenological implications and potential sensitivity was made in Ref.~\cite{Masso:2002np}, where the production of thermal axions above the electroweak scale via quark and gluon interactions was considered. In Refs.~\cite{Graf:2010tv, Salvio:2013iaa} thermal effects were reconsidered, with the finding that the coupling of axions to the top quark had the most significant contribution to the production rate~\cite{Salvio:2013iaa}. Studies below the electroweak scale (although far from the QCDPT) were performed in Refs.~\cite{Ferreira:2018vjj, Arias-Aragon:2020qtn}. Below the QCDPT, hadron processes are relevant, and they were studied in Refs.~\cite{Berezhiani:1992rk, Chang:1993gm, Hannestad:2005df, Kawasaki:2015ofa, Giare:2020vzo, Ferreira:2020bpb}. That is, a hot axion population has been computed for a certain range of temperatures, and thus for a given mass range. A first approach to smoother calculations was made in Ref.~\cite{Arias-Aragon:2020shv}, where the thermal production was calculated across the electroweak phase transition. A further improvement that allows consideration of a wider range of axion coupling to gluons across thresholds was made in Refs.~\cite{DEramo:2021psx, DEramo:2021lgb}. This complete interaction rate has been used in Refs.~\cite{Caloni:2022uya, DEramo:2022nvb, DiValentino:2022edq, Giare:2023aix} to further constrain axion-like particles and axions. Ref.~\cite{Caloni:2022uya} also considered the coupling to photons and made a similar treatment to the one from Ref.~\cite{DEramo:2021lgb}, to smoothly extend the interaction rate to high temperatures. 

Once the axion population in full thermal equilibrium is established, it is expected to decouple due to its feeble interactions while still relativistic.  A relic that decouples in the early Universe while being relativistic can impact, on the one hand, Big Bang Nucleosynthesis (BBN) through the extra contribution to the effective number of neutrinos $N_\nu \equiv 3 + \Delta N_\nu$, where the recent constraints set $N_\nu = 2.880 \pm 0.144$~\cite{Mossa:2020gjc, Yeh:2020mgl}. On the other hand, if they are still relativistic before the photon decoupling epoch, they will leave their imprint on the cosmic microwave
background (CMB) power spectrum. Here, the impact is also parameterized as the effective number of neutrinos $N_{\rm eff} = 3.044 + \DNeff$, measured as $N_{\rm eff}=2.99\pm 0.17$ by Planck 2018~\cite{Planck:2018vyg}. The CMB stage 4 experiment is expected to lower the sensitivity to $\DNeff=0.06$ at 95\%~CL~\cite{CMB-S4:2022ght}.

Besides their impact on the CMB, hot/warm relics impact the Universe's current energy density, affecting the large-scale structure (LSS) spectra. They can potentially cause a reduction in amplitude for small scales in the matter power spectrum because of their free-streaming behavior. The suppression of small-scale matter fluctuations is controlled by the comoving free-streaming length of these relics when they transition from relativistic to nonrelativistic states.

The settlement and evolution of both thermal and non-thermal axions crucially depend upon the cosmological conditions at the moment of and after their generation. An intriguing scenario arises when the Universe undergoes a nonstandard cosmological era preceding BBN~\cite{Allahverdi:2020bys}.
A particularly interesting case corresponds to an early matter dominated (EMD) period, where the Hubble expansion rate was dominated at early times by a species with an energy density that scales as nonrelativistic matter. Such a species could correspond to a long-lived heavy particle (ubiquitous in UV-complete models)~\cite{Vilenkin:1982wt, Coughlan:1983ci, Starobinsky:1994bd, Dine:1995uk, Moroi:1999zb} or to very light primordial black holes that fully evaporate before the onset of BBN. However, examples with more general equation-of-state parameters can occur for oscillating scalar fields with certain potentials. A special case is that of kination in which a ``fast-rolling'' field whose kinetic term dictates the expansion rate of the post-inflation Universe, implying an equation of state $\omega = 1$~\cite{Barrow:1982ei, Ford:1986sy, Spokoiny:1993kt}. Moreover, the transition from an inflaton-dominated to a radiation-dominated Universe, that is, the reheating era, is often assumed to be instantaneous and to occur at a very early time (i.e. a very high temperature). This picture cannot be taken for granted, and scenarios with low-temperature reheating can occur in which the reheating era is prolonged. We refer the reader to Ref.~\cite{Allahverdi:2020bys} for a review on non-standard cosmological scenarios.
Previous studies have extensively explored the NSC scenario within the framework of axion physics. Production of cold DM axions has been studied in different contexts in Refs.~\cite{Steinhardt:1983ia, Lazarides:1990xp, Kawasaki:1995vt, Giudice:2000ex,Visinelli:2009kt, Blinov:2019jqc, Venegas:2021wwm, Arias:2021rer,Arias:2022qjt, Ramberg:2019dgi,Bernal:2021yyb}. Concerning lightweight thermally produced axions, research has predominantly focused on LTR and kination cosmologies, as discussed in Refs.~\cite{Grin:2007yg, Carenza:2021ebx}.

The aim of this work is, on the one hand, to complement and extend the studies done in Refs.~\cite{Grin:2007yg, Carenza:2021ebx}, about the constraints on the production of an axion population in full thermal equilibrium, under the assumption that the Universe had a different expansion period than that due to radiation before BBN. 
On the other hand, we also point out interesting features that have only been briefly discussed in the literature before.

The axion framework with which we will work is the so-called ``hadronic axion models'' that have been extensively used in the literature due to the model-independent axion-gluon coupling, the prime example being the Kim-Shifman-Vainshtein-Zakharov  (KSVZ) model~\cite{kim1979weak, shifman1980can}. In this model, new heavy singlet quarks carry $U(1)_\text{PQ}$ Peccei-Quinn charges, leaving ordinary quarks and leptons without tree-level axion couplings. The most stringent constraint for the KSVZ scenario sets $m_a<0.28$~eV at 95$\%$~CL~\cite{DEramo:2022nvb}, using the complete gluon interaction rate from Ref.~\cite{DEramo:2021psx} and combining the CMB and LSS data. Therefore, to account for the axion-gluon coupling, we also make use of the complete interaction rate from Ref.~\cite{DEramo:2021psx}. We also include in our analysis the coupling to photons, which, even though it is model-dependent, has been extensively exploited for QCD axion searches. This interaction rate has also been extended to high temperatures in Ref.~\cite{Caloni:2022uya}, for hadronic axions. We will treat both interactions independently, assuming that one of them dominates over the other in producing the thermal population.
In this work,
we will impose restrictions on the thermal population in two nonstandard scenarios: low-temperature reheating (LTR) and early matter domination (EMD). The constraints will rely on the results on light massive relics presented in Ref.~\cite{Xu:2021rwg}. Their analysis considers CMB, LSS, and weak-lensing effects. Our findings confirm that the population is dependent on the cosmological history; therefore, it could evade current and future observations. We comment on the thermalization of the population in different cosmologies and about the distinctive free-streaming signatures and their potential to further explore these nonstandard cosmological scenarios. Finally, we comment on the possibility of a coexistence of the axion cold DM population produced by misalignment and the population from the thermal bath produced via freeze-in, which is within the reach of experimental observation. In this work, we do not consider axion generation by other possible mechanisms, such as decays of topological defects. Such contributions can in principle be significant, but their actual value is still in dispute (see Ref.~\cite{ParticleDataGroup:2022pth} and references therein).

The manuscript is organized as follows: In Section~\ref{sec:Couplings} we review the coupling of axions to gluons and photons, focusing on the resultant interaction rates with the thermal bath. In Section~\ref{sec:Thermalaxions} we discuss the thermal axion population produced in a standard radiation-dominated cosmology along with the constraints that we will use. In Section~\ref{sec:NSCThermalaxions} we introduce the features of the NSC to be worked with and address the conditions for thermalization in different cosmological scenarios. Finally, we present our results and analysis for the thermally-produced axion population in two NSC scenarios, LTR and EMD. We point out differences with previous works and comment on their origin. Finally, in Section~\ref{sec:co-existence} we explore the possibility of the co-existence of a hot and cold axion population produced from the bath and the misalignment mechanism, respectively, in a kination-like history. We conclude in Section~\ref{sec:Conclution}.

\section{Relevant axion couplings}
\label{sec:Couplings}
Axions appear as pseudo-Nambu-Goldstone bosons when the so-called Peccei-Quinn (PQ) symmetry is spontaneously broken at an energy scale $f_a$.
Assuming $f_a$ is higher than the electroweak scale (the so-called invisible axion model), axion scattering in the early Universe may lead to a dark population, either in or out of thermal equilibrium with the SM bath. For temperatures smaller than $f_a$, the effective Lagrangian density for the axion $a$ contains couplings to gluons and photons and is given by
\be
    \mathcal L= \frac12 \left(\partial_\mu a\right) \left(\partial^\mu a\right) + \mathcal L_{ag} + \mathcal L_{a\gamma}\,.
\ee

A distinctive feature of the axion is that its mass has a definite relationship with the PQ scale through the topological susceptibility of QCD, which has been evaluated in the chiral limit~\cite{Crewther:1977ce, DiVecchia:1980yfw}, NNLO in chiral perturbation theory~\cite{Gorghetto:2018ocs}, and directly via QCD lattice simulations~\cite{Borsanyi:2016ksw}, giving 
\be
    \ma \simeq 5.69~\mbox{meV} \left(\frac{10^9~\mbox{GeV}}{f_a} \right),
\ee
where $\ma$ corresponds to the mass of the axion at zero temperature.

\subsection{Coupling to gluons}
\label{sec:CouplingGluon}
Let us first examine the coupling to gluons, written as
\be
    \mathcal L_{ag} \subset \frac{\alpha_s}{8\pi\, f_a} \, a\, G_{\mu\nu}^i\, \tilde G^{\mu\nu, i},
\ee
where $\alpha_s$ is the strong coupling and $G$ and $\tilde G$ are the gluon field strength and its dual, respectively.
For temperatures above the QCDPT $T\gg \Tqcd$, the main channels that can thermalize an axion population are the 2-to-2 scattering processes $g + g \to g + a$, $q + \bar q \to g + a$, $q + g \to q + a$ and $\bar q + g \to \bar q + a$, with $g$ being a gluon, $q$ a quark and $\bar q$ an antiquark.  The total axion production rate was obtained in Refs.~\cite{Masso:2002np, Graf:2010tv} by a hard thermal loop approximation. An improvement including higher-order effects was made consistently in Ref.~\cite{Salvio:2013iaa}, where couplings to quarks were also included. In general, the interaction rate density of axions with gluons $\gamma_{gg}(T)$ is proportional to
\be
    \gamma_{gg}(T) \propto \frac{\zeta(3)}{(2\pi)^5}\, \frac{T^6}{f_a^2}\,.
\ee
As the Universe evolves and the temperature decreases, quarks hadronize, and therefore other processes become responsible for thermalization. For temperatures $T\ll \Tqcd$, the main channels are $a + \pi \to \pi + \pi$ and $a + N \to N + \pi$, where $\pi$ stands for $\pi^0$ and $\pi^\pm$. Since pions are more abundant than nucleons, the former process dominates over the latter. An analytical expression for such an interaction rate is \cite{Hannestad:2005df}
\begin{equation}
    \Gamma_{\pi\pi}(T) \equiv \frac{\gamma_{\pi\pi}(T)}{n_a^{\rm eq}(T)} =  A\, C_{a\pi}^2\, h\left(\frac{m_\pi}{T}\right) \frac{T^5}{\left(f_a\, f_\pi\right)^2}\,,
    \label{eq:rate}
\end{equation}
where $m_\pi$ and $f_\pi$ are the mass and the coupling constant of the pion, respectively, $A \simeq 0.215$, and  $h(m_\pi/T)$ is a monotonically decreasing function for $m_\pi/T > 1$, as $h(0) = 1$~\cite{Hannestad:2005df}. $C_{a\pi}$ is the dimensionless axion-pion coupling constant given by
\begin{equation}
    C_{a\pi} \equiv \frac{1-r}{3\, (1+r)}\,,
\end{equation}
where $r\equiv m_u/m_d \simeq 0.56$ is the ratio of up- and down-quark masses. The validity of this expression has been questioned in Ref.~\cite{DiLuzio:2021vjd} for decoupling temperatures above $\sim 62$~MeV, where the effective field theory is expected to break down. Instead, a unitarized next-to-leading-order chiral perturbation theory can be used, extending the reliability of the interaction rate to decoupling temperatures up to $\sim 155$~MeV~\cite{DiLuzio:2022gsc}. Alternatively, the axion-pion rate can be extracted directly from the experimental data on pion scattering, by rescaling the corresponding cross sections~\cite{Notari:2022zxo}.
 
\begin{figure}[t]
    \centering
    \begin{subfigure}[b]{0.49\textwidth}
    \centering
    \includegraphics[width=1\textwidth]{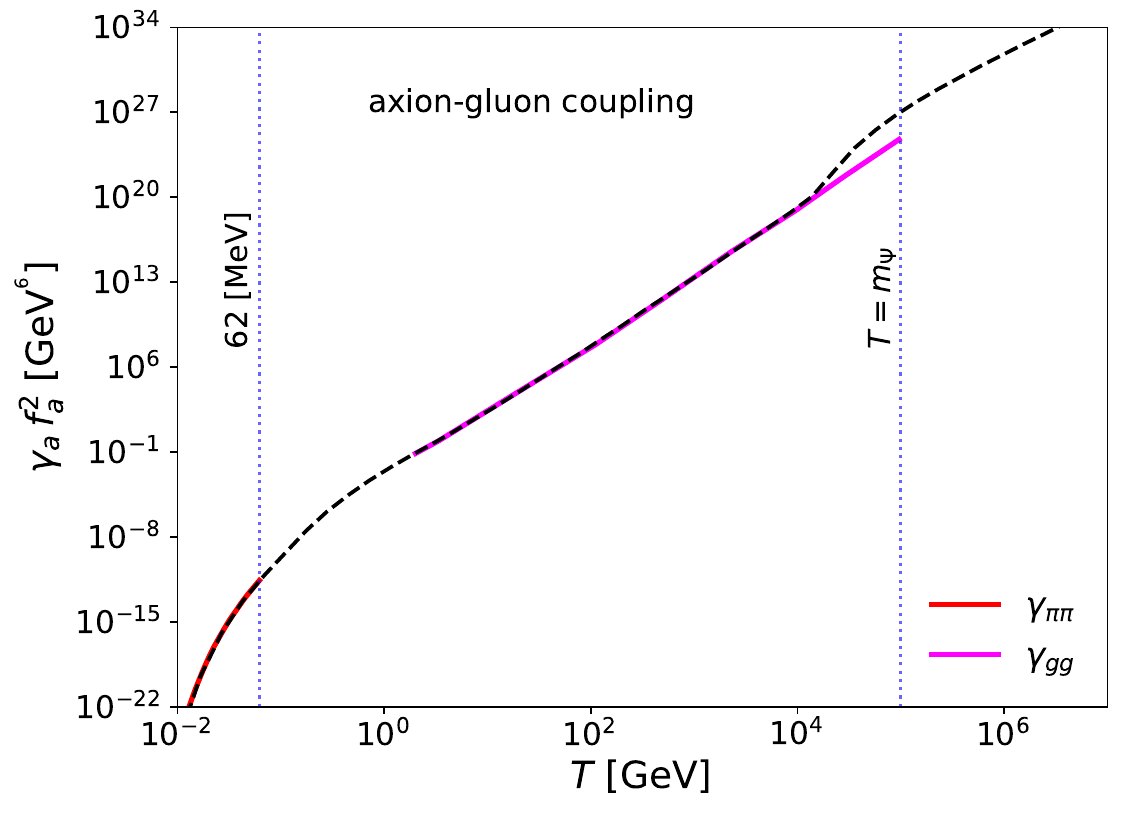}
    \caption{}
    \label{fig:smooth_rate}
    \end{subfigure}
    \hfill
    \begin{subfigure}[b]{0.49\textwidth}
    \centering
    \includegraphics[width=1\textwidth]{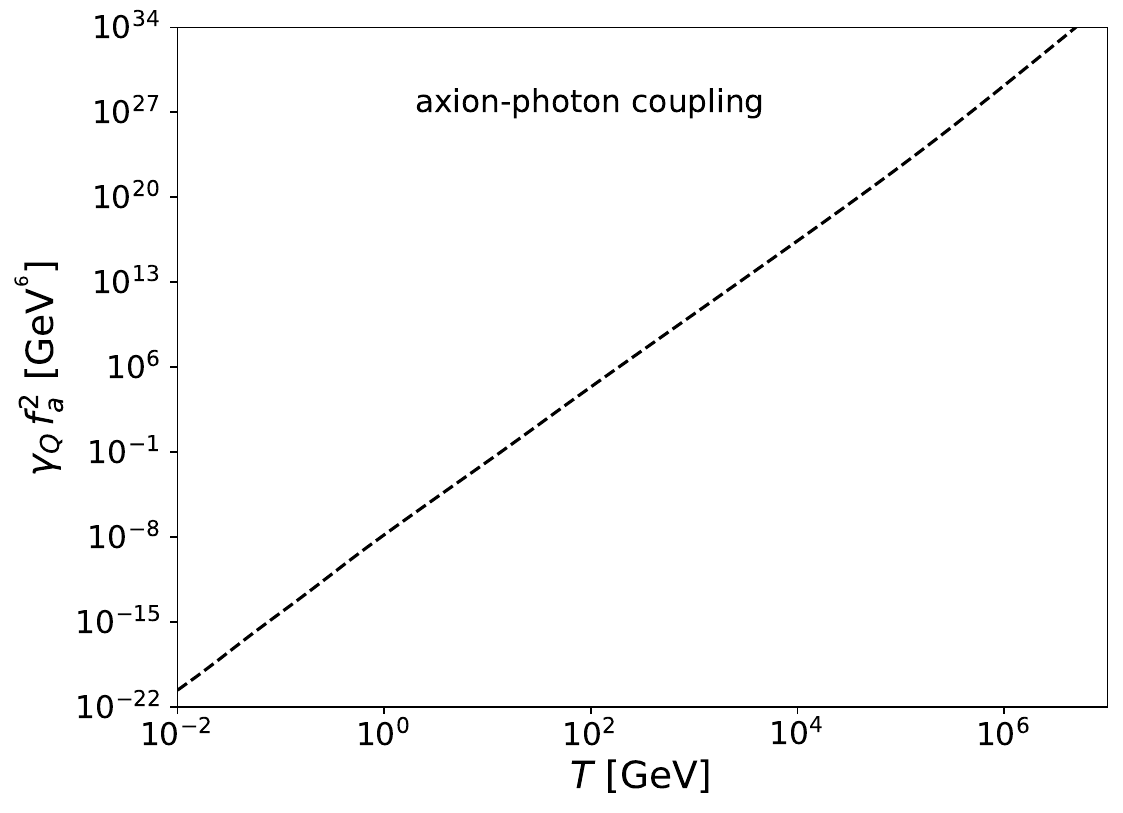}
    \caption{}
    \label{fig:smooth_rate2}
    \end{subfigure}
   \caption{(a) KSVZ total interaction rate of axions across both the heavy colored Peccei-Quinn fermion mass $m_\Psi=10^5$~GeV and the QCDPT, represented by the dashed black line (adapted from Ref.~\cite{DEramo:2021lgb}). Solid red and magenta lines show the pion and gluon scattering rates quoted in the text. (b) Interaction rate with photons (Primakoff effect).} 
\end{figure}

A step forward in accounting for this axion rate was made in Refs.~\cite{DEramo:2021psx, DEramo:2021lgb}, where a smooth transition of the interaction rate of axions to gluons across the QCDPT was achieved, by computing it above and below the confinement scale and interpolating between the two regimes.  
They also treat the threshold at the mass of the heavy PQ fermion \(\Psi\), above which the binary collisions of \(\Psi\) become the dominant production channel for the axions. Their result and the corresponding comparison with the interaction rates mentioned above can be seen in Fig.~\ref{fig:smooth_rate} for the KSVZ model. The vertical dotted blue lines correspond to $T = 62$~MeV and $T = m_\Psi$. In the following, we will use the continuous smooth rate of Ref.~\cite{DEramo:2021psx} for convenience. To account for the importance of using the continuous rate across the QCDPT we present in Appendix~\ref{sec:appendix} a comparison of the bounds presented in this work with those that would be obtained if only the axion-pion interaction was used; cf. Eq.~\eqref{eq:rate}. Although the interpolated rate suffers from greater uncertainties, we have explicitly checked that using the rates from Refs.~\cite{DiLuzio:2022gsc, Notari:2022zxo} does not significantly alter the limits nor the features presented in the next sections.

\subsection{Coupling to photons}
The coupling of axions to photons is undoubtedly the most exploited one to search for these particles, even though the strength is model dependent. It reads
\be
    \mathcal L_{a\gamma\gamma} = -\frac{g_{a\gamma\gamma}}{4}\, a\, F_{\mu\nu}\, \tilde F^{\mu\nu},
\ee
where $F$ is the electromagnetic field tensor and $\tilde F$ its dual. The coupling constant $g_{a\gamma\gamma}$ is given by
\be
    g_{a\gamma\gamma} = \frac{\alpha_{\rm em}}{2\pi\, f_a} \left(\frac{E}{N} - 1.92\right) \equiv \frac{\tilde g_\gamma}{f_a}\,.
\ee
Here $\alpha_{\rm em}$ is the fine-structure constant, while $E$ and $N$ are the electromagnetic and color anomalies of the axial current associated with the axion, respectively. There are two benchmark axion models. In a model known as the Dine-Fishler-Srednicki-Zhitnitsky (DFSZ) model~\cite{Zhitnitsky:1980tq, Dine:1981rt} at least two additional Higgs doublets are needed, and ordinary quarks and leptons carry PQ charges. The KSVZ model\cite{kim1979weak, shifman1980can} instead features new heavy singlet quarks carrying $U(1)_{\rm PQ}$ charges, leaving normal quarks and leptons without tree-level couplings to axions. For the KSVZ models $E/N = 0$, while for the DFSZ models, $E/N = 8/3$. There are, however, extensions to these models, in particular to the KSVZ case, that can spawn several other values of $E/N$ (see, e.g., Ref.~\cite{Irastorza:2018dyq}).
In this work, we assume that the axion comes from a KSVZ-like model and therefore does not couple to SM fermions directly.

The axion-photon coupling contributes to the formation of a thermal population of axions via the Primakoff effect. The rate of axion production due to scattering in a multicomponent charged plasma is given by~\cite{Bolz:2000fu, Cadamuro:2011fd}
\be \label{eq:GammaPrimakoff}
    \Gamma_Q \simeq \frac{\alpha_{\rm em}\, g_{a\gamma\gamma}^2\, \pi^2}{36\, \zeta(3)} \left[\ln\left(\frac{T^2}{m_{\gamma}^2}\right) + 0.82\right] n_Q\,,
\ee
where $n_Q \equiv \sum_i Q_i\, n_i \equiv \left(\zeta(3)/\pi^2\right) g_Q(T)\, T^3$ is the effective number density of charged particles, with $Q_i$ the charge of the $i^\text{th}$ particle species. $g_Q(T)$ accounts for the effective number of charged relativistic degrees of freedom. In the hot early Universe, the photon has an effective mass (plasmon mass) given by $m_\gamma(T) = g_Q^{1/2}\, T/(6\, \alpha_{\rm em})$. 
Fig.~\ref{fig:smooth_rate2} shows the interaction rate density defined as $\gamma_Q \equiv n_a^\text{eq}\, \Gamma_Q(T)$.\\

Before closing this section, we note that the coupling of axions to photons allows for axion decay, with a rate given by~\cite{ParticleDataGroup:2022pth}
\be
    \Gamma_{a\to \gamma \gamma } = \frac{g_{a\gamma\gamma}\, m_a^3}{64 \pi} \simeq 1.1 \times 10^{-24}\, \mbox{s}^{-1} \left(\frac{m_a}{\mbox{eV}}\right)^5,
\ee
which implies that axions above $m_a\gtrsim 20$~eV have decayed before the present time. 

From the above analysis, it can be seen that at very high temperatures, above the new heavy fermion mass, the Primakoff interaction is efficient in thermalizing axion interactions, but below that scale, processes involving the gluon coupling take over up to temperatures similar to the pion mass, where the interaction ceases. Thus, for masses below the eV scale, in a radiation-dominated Universe, the gluon coupling is responsible for keeping thermal equilibrium to lower temperatures.

\section{Thermal axions in standard cosmology} \label{sec:Thermalaxions}

\subsection{The population of thermal axions}
\begin{figure}[t]
    \centering
    \begin{subfigure}[b]{0.49\textwidth}
    \centering    \includegraphics[width=1\textwidth]{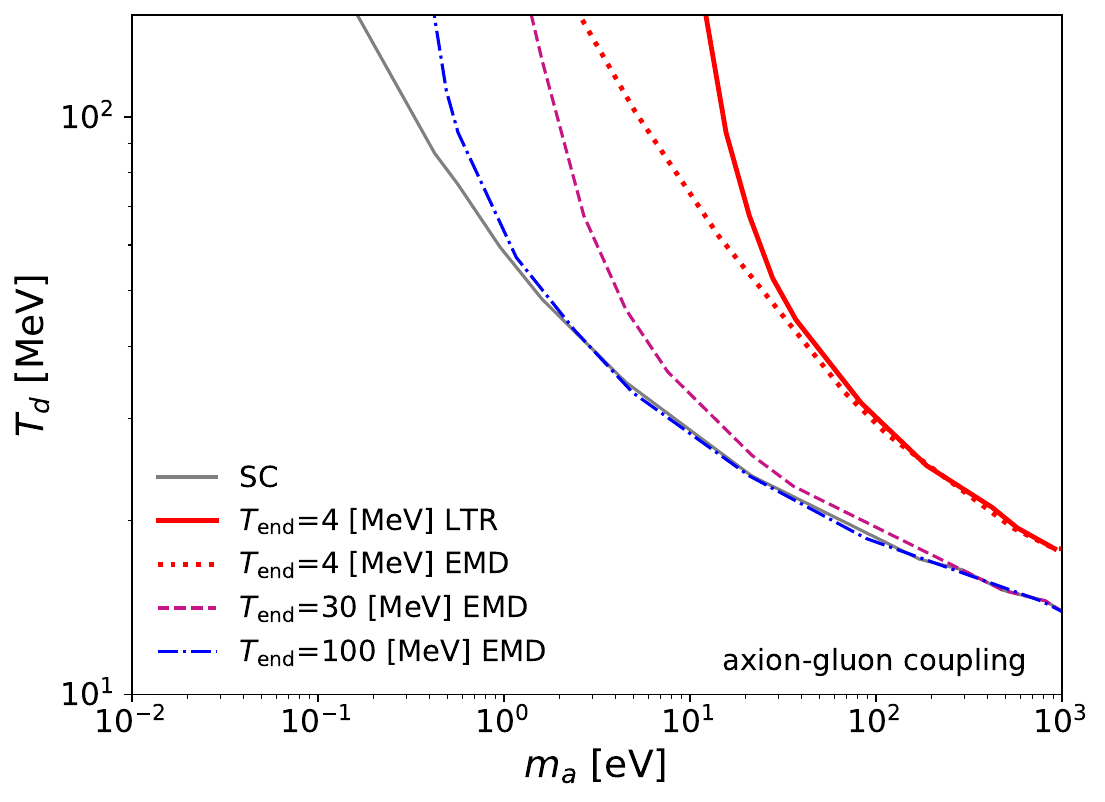}
    \caption{}
    \end{subfigure}
    \hfill
    \begin{subfigure}[b]{0.5\textwidth}
    \centering
    \includegraphics[width=1\textwidth]{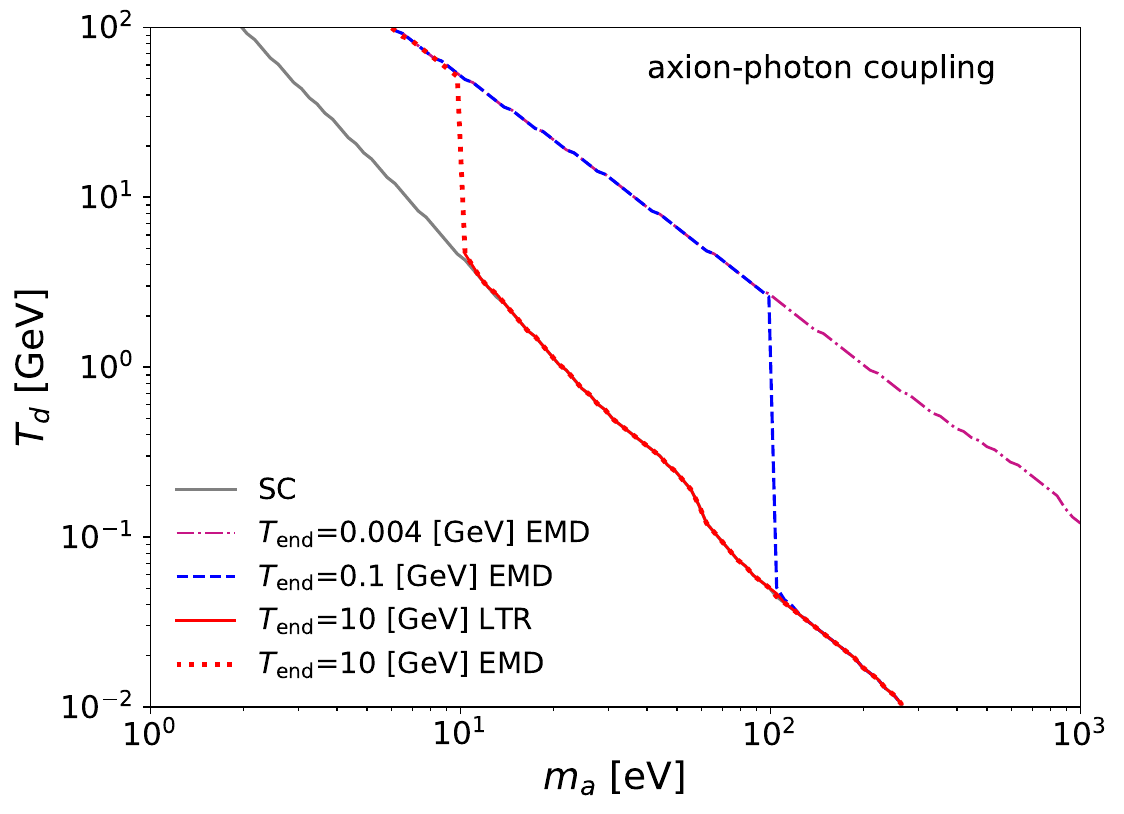}
    \caption{}
    \label{}
    \end{subfigure}
     \caption{  Decoupling temperature as a function of axion mass considering the coupling of axions to (a) gluons and (b) photons. Gray lines correspond to the standard cosmology, whereas solid lines are for LTR  and segmented colored lines correspond to different EMD scenarios.
     }
     \label{fig:decouplingT}
\end{figure}

If the interaction rate with the SM particles is strong enough, axions reach chemical equilibrium with the primordial plasma in the early Universe\footnote{We comment on the thermalization for the SC and its subtleties in Section~\ref{sec:NSCThermalaxions}.}. However, when the interaction rate becomes smaller than the expansion rate of the Universe, axions decouple from the thermal bath and freeze out. An estimation of the decoupling temperature $\Td$ can be obtained from 
\begin{equation} \label{eq:approx}
     \Gamma (\Td) = H_R(\Td)\,,
\end{equation}
where $H_R$ is the Hubble expansion rate in a radiation-dominated Universe,
\begin{equation} \label{eq:H(T)rad}
    H_R(T) \equiv \sqrt{\frac{\rR(T)}{3\, M_P^2}} = \frac{\pi}{3}\, \sqrt{\frac{\gs(T)}{10}}\, \frac{T^2}{M_P}\,,
\end{equation}
where the SM radiation energy density is
\begin{equation}
    \rR(T) \equiv \frac{\pi^2}{30}\, \gs(T)\, T^4,
\end{equation}
with $M_P \simeq 2.4 \times 10^{-18}$~GeV being the reduced Planck mass and $\gs(T)$ the number of relativistic degrees of freedom contributing to $\rR$~\cite{Drees:2015exa}.
The decoupling temperature as a function of the axion mass is shown in Fig.~\ref{fig:decouplingT} with solid gray lines in standard cosmology for axion couplings to gluons (left) and photons (right). The rapid change in slope when $\ma \simeq \mathcal{O}\left(10^{-2}\right)$~eV corresponds to a decoupling temperature around the QCDPT.
For the interaction with photons, the decoupling temperature can be found analytically to be
\be
    T_\text{d}^P = 91\, \frac{\sqrt{\gs(T_\text{d}^P)}}{g_Q(T_\text{d}^P)} \left(\frac{f_a}{10^6\,\mbox{GeV}}\right)^2 \left(\frac{\tilde g_\gamma}{\alpha_{\rm em}/2\pi}\right) \mbox{GeV},
\ee
where $\gs(T)$ corresponds to the number of relativistic degrees of freedom contributing to the SM energy density.

At the moment of their decoupling,\footnote{We assume instantaneous decoupling throughout this work.} axions have the same temperature as the SM bath, and therefore their number and energy densities are given by
\begin{align}
    n_a(\Td) &= \frac{\zeta(3)}{\pi^2}\, \Td^3,\\
    \rho_a(\Td) &= \frac{\pi^2}{30}\, \Td^4,
\end{align}
since they decouple while being relativistic.
As their total number after decoupling is conserved, their number density today is
\be
    n_a(T_0) = \frac{\zeta(3)}{\pi^2}\, \frac{\gss(T_0)}{\gss(\Td)}\, T_0^3 \simeq 80 \left(\frac{10}{\gss(\Td)}\right) \mbox{cm}^{-3},
\ee
where the conservation of the SM entropy in the standard cosmological scenario was used, $\gss(T)$ is the number of relativistic degrees of freedom contributing to the SM entropy~\cite{Drees:2015exa}, and $T_0$ is the photon temperature today. If axions become nonrelativistic before today, their energy density is simply $\rho_a(T_0) = m_a\, n_a(T_0)$, 
therefore, their contribution to the total energy density of the Universe is
\begin{equation}
     \Omega_a h^2=0.007\left(\frac{m_a}{ \mbox{eV}}\right)\left(\frac{10}{\gss(\Td)}\right).
     \label{eq:sc_relic}
 \end{equation}
Equivalently, it can also be expressed in terms of today's axion temperature $T_{a,0}$,
\be
    \Omega_a h^2=0.02 \left(\frac{m_a}{\mbox{eV}}\right)\left(\frac{T_{a,0}}{T_{0}}\right)^3,
    \label{eq:axion_relic}
\ee
by noticing that after their decoupling, the axions leave thermal equilibrium, and their temperature $T_a$ redshifts until today in terms of the photon temperature as\footnote{After decoupling, axions are not in thermal equilibrium and therefore they do not follow a thermal distribution anymore. There, the {\it effective} temperature $T_a$ must be understood as a proxy of their momentum.}
\be
     T_a(R) = \Td\, \frac{R_{\rm d}}{R} = T \left(\frac{\gss(T)}{\gss(\Td)}\right)^{1/3},
     \label{eq:Td_sc}
\ee
where $R$ is the cosmic scale factor and $R_\text{d} \equiv R(\Td)$.

Alternatively, if axions remain relativistic until today, their energy density at present is
\begin{equation}
    \Omega_a h^2 = \frac{\pi^2}{30} \left(\frac{\gss(T_0)}{\gss(\Td)}\right)^\frac43\, \frac{T_0^4}{\rho_c}\, h^2 \simeq 4 \times 10^{-6} \left(\frac{10}{\gss(\Td)}\right)^{4/3}.
    \label{eq:sc_relic_rel}
\end{equation}
By equating Eqs.~\eqref{eq:sc_relic} and~\eqref{eq:sc_relic_rel}, we find that the transition between axions being relativistic and nonrelativistic occurs at a mass of 
\begin{equation}
    m_a \simeq \frac{\pi^4}{30\, \zeta(3)} \left(\frac{\gss(T_0)}{\gss(\Td)}\right)^\frac13 T_0 \simeq 0.4 \left(\frac{10}{\gss(\Td)}\right)^\frac13 \text{meV},
\end{equation}
which shows that axions of masses roughly below $10^{-3}$~eV are still relativistic today. Due to their early detachment from the thermal bath, particles with masses below the meV undergo significant entropy dilution from the SM, resulting in a negligible impact on the relic abundance and, consequently, on $\Delta N_{\rm eff}$, as discussed in the subsequent subsection. This suppression is further pronounced during non-adiabatic expansion because of entropy dilution. Hence, the mass target for this study is set above the meV scale.

\subsection{Constraints on thermal axions}
In recent years, there has been extensive effort focused on constraining light, massive relics beyond the SM by integrating observations across various cosmic epochs. One approach has been to combine data from the early Universe, such as the Cosmic Microwave Background (CMB), with measurements from the late Universe, including galaxy clustering and gravitational lensing~\cite{Baumann:2015rya, Banerjee:2016suz}.
This integration has attracted increased attention due to some preliminary observations suggesting the potential necessity of a certain amount of warm DM to address existing tensions, such as the $H_0$ and $\sigma_8$ tensions~\cite{Verde:2019ivm, Battye:2014qga}. 

Restrictions on the population of any thermal relic are usually separated depending on when in the cosmological timeline they transitioned to become nonrelativistic. Let us estimate this transition by comparing the momentum of the axions to their mass
\be
    \langle p_a\rangle \simeq 2.7\, T_{a, \text{nr}} \sim m_a\,,
\ee
where $T_{a, \text{nr}}$ is the temperature of the axion at the moment it becomes non-relativistic. As the temperature of the axion only redshifts, we can trade the temperature $T_{a,\text{nr}}$ for the corresponding scale factor at that time, $T_{a,\text{nr}} = T_{a,0}/R_\text{nr}$, and therefore
\be
    \frac{1}{R_\text{nr}} = 1 + z_\text{nr} \simeq \frac{m_a}{2.7\, T_0} \left(\frac{\gss(\Td)}{\gss(T_0)}\right)^{1/3} \simeq 3436\, \frac{m_a}{1.4~\mbox{eV}} \left(\frac{\gss(\Td)}{14.5}\right)^{1/3}.
    \label{eq:nr_st}
\ee

For coupling to gluons, axions of mass $m_a \gtrsim 1.4$~eV turn non-relativistic around the moment of matter-radiation equality, which corresponds to a decoupling temperature of around $\Td \lesssim 50$~MeV (see Fig.~\ref{fig:decouplingT}). Instead, for the Primakoff interaction, this happens for axions of $m_a \gtrsim 0.8$~eV, that decouple at a temperature $\Td \lesssim 400$~GeV. 
Relativistic axions at matter-radiation equality contribute to the number of effective neutrinos present at the CMB decoupling. On the other hand, for those that transition already into a nonrelativistic state, their velocity dispersion can still be high enough such that there is an effective free-streaming scale below which perturbations in the relic component are suppressed. If they constitute a significant fraction of the energy budget, this suppression hinders the growth of matter perturbations on small scales, and they can still be constrained. 

\subsubsection*{Hot axions}
The relativistic and semi-relativistic populations of axions at CMB decoupling behave as dark radiation and contribute to the so-called effective number of neutrinos at that time, which can be accounted for using
\be
    \rR = \rho_\gamma\left[ 1 + \frac78 \left(\frac{4}{11}\right)^{4/3} N_{\rm eff}\right],
\ee
where the contribution from a dark relic is usually parametrized as
\be
    N_{\rm eff} \equiv N_{\rm eff}^{\rm SM} + \DNeff\,,
\ee
with
\be
    \DNeff = \left.\frac87 \left(\frac{11}{4}\right)^{4/3} \frac{\rho_{a}}{\rho_\gamma}\right\vert_{\rm CMB}.
\ee
In the case of a relativistic relic that was once in full thermal equilibrium, its contribution can be written as
\be
    \DNeff =  \left(\frac47\right) \left(\frac{11}{4}\right)^{4/3}\left(\frac{\gss(T_{\rm CMB})}{\gss(\Td)}\right)^{4/3}\simeq 2.1 \left(\frac{T_{a,0}}{T_{0}}\right)^4.
    \label{eq:massless_Neff}
\ee
    
Thus, for axions that decoupled well before the QCDPT, the number of relativistic degrees of freedom reaches a plateau at $\gss(\Td)=106.8$, and their contribution to the effective number of neutrinos is always $\DNeff\sim 0.027$. The recent value of $N_{\rm eff}=2.99\pm 0.17$ has been found, with a 95\% CL upper limit from Planck 2018 of $\DNeff\lesssim 0.35$~\cite{Planck:2018vyg, Cielo:2023bqp}. In Ref.~\cite{DEramo:2022nvb}, Planck temperature and polarization data was used to set the bound $\Delta N_{\rm eff} < 0.3$, which restricts the mass of the axion to $m_a < 1.04$~eV. On the contrary, by introducing LSS data from Planck lensing and BAO (6dFGS, SDSS MGS and BOSS DR12 surveys) the bound drops to $\Delta N_{\rm eff}<0.23$ yielding $m_a<0.28$~eV, reinforcing the strength of the LSS data to constrain light relics.
Axions lighter than $\sim 0.1~$eV escape detection because they decouple above the QCDPT and, therefore, their abundance is strongly diminished. The coupling to two photons, on the other hand, is not effective enough to place a constraint on relativistic axions during the CMB decoupling epoch. 

Upcoming CMB experiments, such as SPT-3G~\cite{SPT-3G:2014dbx} and the Simons Observatory~\cite{SimonsObservatory:2018koc}, will soon improve Planck's precision in $N_\text{eff}$. In particular, CMB-S4~\cite{Abazajian:2019eic} and CMB-HD~\cite{CMB-HD:2022bsz} will be sensitive to a precision of $\DNeff \sim 0.06$ and $\DNeff \sim 0.027$ at 95\% CL, respectively. As calculated in Ref.~\cite{Yeh:2022heq}, a combined analysis from BBN and CMB results in $N_\text{eff} = 2.880 \pm 0.144$. The next generation of satellite missions, such as COrE~\cite{COrE:2011bfs} and Euclid~\cite{EUCLID:2011zbd}, shall impose limits at $2\sigma$ on $\DNeff \lesssim 0.013$. Furthermore, as mentioned in Ref.~\cite{Ben-Dayan:2019gll}, a hypothetical cosmic-variance-limited CMB polarization experiment could presumably reduce the limit to as low as $\DNeff \lesssim 3 \times 10^{-6}$, although this seems experimentally challenging.

\subsubsection*{Warm axions}
A second restriction on the mass of hot relics comes from their impact on the total energy density of the Universe today. The hot massive relic components in the Universe are included in the neutrino density of the Universe, which is given by $\Omega_\nu = \sum {m_\nu}/({93.14 \,h^2~\text{eV}})$. Planck 2018+lensing+BAO data have set a stringent limit of $\sum m_\nu < 0.12$~eV. Subtracting the contribution of active neutrinos considering the minimal mass allowed by neutrino flavor oscillation experiments  $\sum m_\nu> 0.06$~eV~\cite{Planck:2018vyg, ParticleDataGroup:2022pth}, we obtain that the contribution from a different kind of hot relic must satisfy
\be
    \Omega_h h^2 < 2 \times 10^{-3}.
\ee
From Eq.~\eqref{eq:sc_relic} this yields $m_a < 0.35$~eV.

The hot DM component streams away from structures below its free streaming length, reducing the growth of matter fluctuations at smaller distances. The size of the suppression depends on the present hot DM abundance~\cite{Ali-Haimoud:2012fzp}.
The free-streaming length $\lambda_{\rm fs}$ (sometimes called the free-streaming horizon) gives the distance traveled by the hot particle at a given time. Physically, this sets the scale below which collisionless particles can not stay confined in gravitational potential wells. It is defined by
\be
    \lambda_{\rm fs}(t) = \int_{t_0}^t \frac{v(t')}{R(t')}\, dt'.
\ee
While axions are relativistic, they travel at the speed of light and their
free-streaming length is simply equal to the Hubble radius. When they become non-relativistic, their thermal velocity drops to
\be
\langle v_a\rangle = \frac{\langle p_a\rangle}{m_a} \simeq 2.7\, \frac{T_{a}}{m_a} \simeq 188 \left(\frac{\mbox{eV}}{m_a}\right) (1+z)\,\frac{T_{a,0}}{T_{0}}\,\mbox{km\, s}^{-1}
    \label{eq:velocity}
\ee
for $z < z_{\rm nr}$.
In standard cosmology, sub-eV axions become non-relativistic in the matter-dominated era. Their free-streaming length peaks around the moment they become non-relativistic and then reaches a steady state. Therefore, it is customary to approximate $\lambda_{\rm fs} \simeq \lambda_{\rm fs}^{\rm nr}\sim (R_{\rm nr}\, H_{\rm nr})^{-1}$ which is given by
\be
    \lambda_{\rm fs}^{\rm nr} \simeq 446 \left(\frac{\mbox{eV}}{m_a}\right)^{1/2} \left(\frac{T_{a,0}}{T_0}\right)^{1/2}\, h^{-1}\, \mbox{Mpc}.
\ee
Axions with larger masses ($\gtrsim$ eV), will become non-relativistic before the radiation-matter transition, assuming a standard cosmology. Integrating the above for $t>t_{\rm eq}>t_{\rm nr}$ we obtain
\be
    \lambda_{\rm fs} = 2\, \frac{t_{\rm nr}}{R_{\rm nr}} \left[1 + \frac12 \log\left({\frac{t_{\rm eq}}{t_{\rm nr}}}\right) \right] + 3\, \frac{R_{\rm nr}\, t_{\rm eq}}{R_{\rm eq}^2}\left[1-\left(\frac{t_{\rm eq}}{t}\right)^{1/3}\right].
\ee
The maximum of this expression occurs around $t \simeq t_{\rm eq}$, and then it reaches a steady value. Using Eq.~(\ref{eq:nr_st}), the free-streaming at the moment of matter-radiation equality, in terms of the mass of the particle and today's temperature (related to today's thermal velocity) can be written as
\begin{align}
    \lfs \simeq \frac{113\, \text{Mpc}}{m_{a}/\rm{eV}} \left(\frac{T_{a,0}}{T_{0}}\right) \left[1 + \ln\left(0.42\, \frac{m_{a}}{\mbox{eV}} \left(\frac{T_{0}}{T_{a,0}}\right)\right) \right].
\end{align}
The free-streaming for particles in this mass range is important up to the point when the particles are indistinguishable from a cold DM relic and their free-streaming distance is negligible.

\subsubsection*{Cold axions}
Thermal axions will behave as a cold relic at matter-radiation equality -- thus contributing to the cold DM energy density -- roughly when their velocity satisfies $\left.\langle v_a \rangle\right|_{z_\text{eq}} \ll 1$, i.e. for masses well above the eV scale.  They could overclose the Universe if their relic density is higher than the cold DM one, $\Omega_c\, h^2\sim 0.12$. Assuming a standard cosmological history, this occurs for the axion-gluon coupling, for masses of $m_a\gtrsim 15~$eV, with a free-streaming length of $\lambda_{\rm fs}\lesssim 44$~Mpc. For the photon coupling, the bound saturates at $m_a\gtrsim 54$~eV, with $\lambda_{\rm fs}\lesssim 12$~Mpc.
\subsubsection*{Constraints}
To study the phenomenology of the formation of thermal axions, we will assume each interaction acts independently of the other. To derive constraints, we make use of the results presented in Fig.~2 of Ref.~\cite{Xu:2021rwg} for Weyl fermions, 
which are based on cosmic microwave background (CMB),  large-scale structure (LSS) and weak-lensing data. The analysis incorporates the full likelihoods of TT, TE, EE, low E, and lensing from Planck 2018 data~\cite{Planck:2018vyg}. Additionally, weak-lensing data from the Canada-France-Hawaii Telescope (CFHTLens) collaboration~\cite{Heymans:2013fya} are included, consisting of 2-point correlation functions of galaxy ellipticities. Furthermore, galaxy power-spectrum data from the Baryon Oscillation Spectroscopic Survey (BOSS) data release 12~\cite{BOSS:2016wmc} are incorporated. This dataset contains spectroscopic information from a large number of galaxies, divided into two redshift bins ($z=0.38$ and $z=0.61$). The analysis focuses on the linear regime, up to a maximum wave number of $k_{\rm max}=0.25\, h\, \rm Mpc^{-1}$, which is within the validity range of the perturbative approach. The free streaming of light relics on small scales suppresses the growth of CDM+baryon fluctuations at the linear level, with a scale-dependent suppression characterized by $k_{\rm fs}=2\pi/\lambda_{\rm fs}$, where the amplitude is set by the relic abundance, $\Omega h^2$. The analysis of Ref.~\cite{Xu:2021rwg} is distinguished from previous works on light massive relics in two key aspects: first, the inclusion of the full-shape galaxy data shows that they improve their constraints concerning the usual CMB+BOSS-BAO by approximately 30$\%$. Second, the inclusion of weak-lensing data breaks the degeneracy between the CDM and the thermal relic abundance that would otherwise be present. That is, weak-lensing data prefer a fixed value of CDM (close to the $\Lambda$CDM one),  pushing the mass of the light relic to smaller values.

In their analysis, it is assumed that the relic was once in full thermal equilibrium with the primordial bath. A standard cosmological history is implicitly assumed, and a prior is set with the lowest possible temperature for a scalar relic at 0.91~K. The constraints obtained are currently the tightest, with lower mass limits of $m_X \geq 2.3$, 11, 1.1, 1.6~eV for Weyl fermions, scalars, vectors, and Dirac fermion relics, respectively.

\begin{figure}[t]
    \centering
    \includegraphics[scale=0.5]{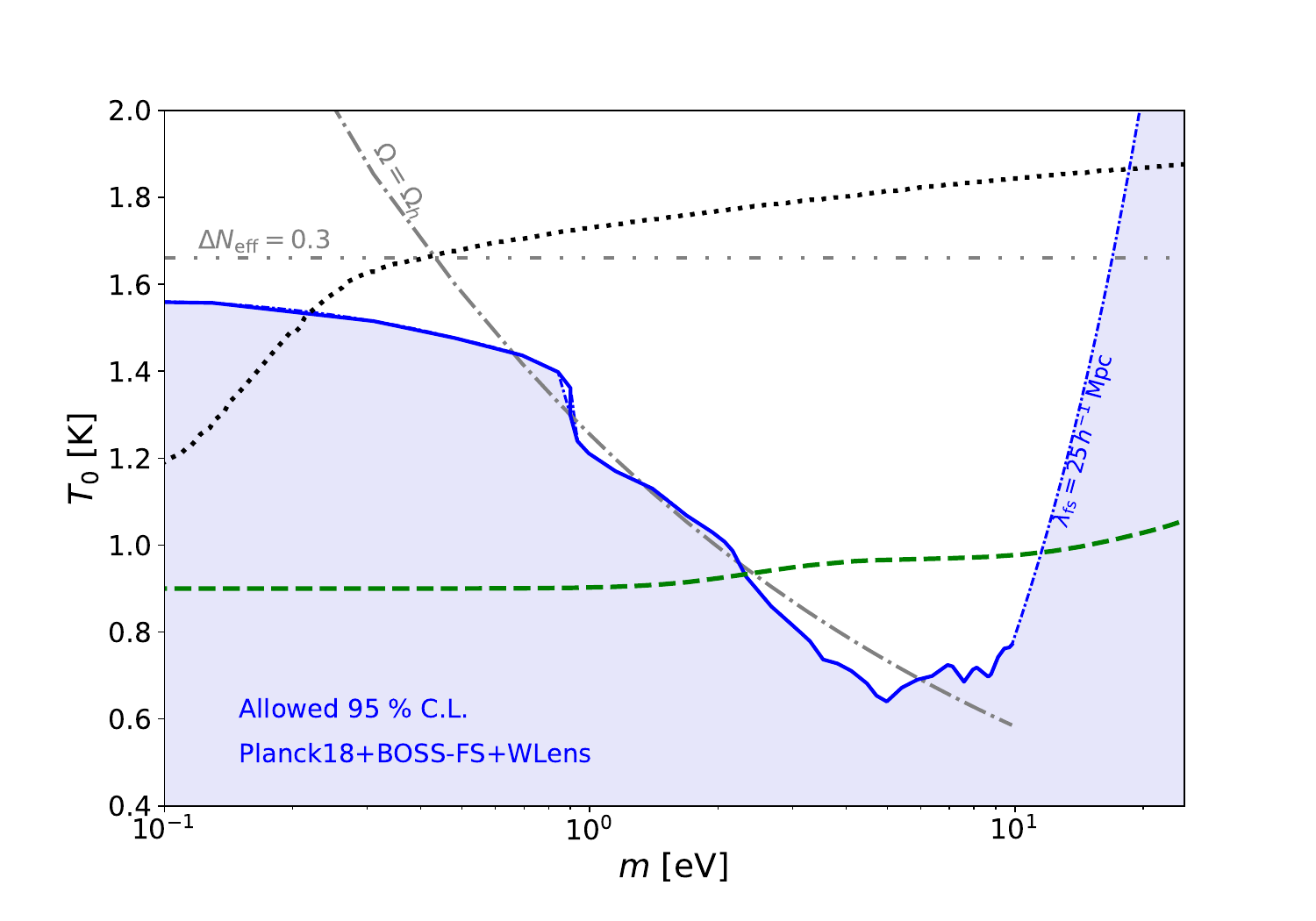}
    \caption{The blue region corresponds to the 95 $\%$~CL allowed parameter space for light massive scalar relics obtained in Ref.~\cite{Xu:2021rwg}, and the white region is excluded. The black dotted line represents the axion temperature today as a function of the mass for the QCD axion population produced in standard cosmology via their interaction with gluons. The dashed green line is the analog for the coupling to photons. Above $m \sim 10$~eV we have assumed an asymptotic value for the free-streaming length of $\lambda_{\rm fs}=25~h^{-1}$~Mpc; see the text for details.}
    \label{fig:bounds_22}
\end{figure}

To map the constraints from Weyl fermions to axions, it is assumed that both behave indistinguishably (at least on linear scales) in terms of relic density, contribution to $\DNeff$, and today's average velocity. This is ensured by the following mappings:
\be
    m_a \simeq 0.986\, m_W \qquad \text{ and} \qquad T_a \simeq 1.15\, T_W\,,
\ee
where $m_W$ and $T_W$ are the mass and temperature of the Weyl fermion thermal relic. All in all, the results of Ref.~\cite{Xu:2021rwg} for light massive scalars once in full thermal equilibrium -- referred to as Planck18+BOSS-FS+WLens -- are depicted in Fig.~\ref{fig:bounds_22}. Data support everything within the blue region at a 95$\%$~CL.

Ref.~\cite{Xu:2021rwg} analyzed masses up to $\sim 10$~eV, since higher masses lead to a free-streaming scale beyond the validity of the linear regime ($k_{\rm max} = 0.25~h$~Mpc$^{-1}$). To connect with the results of previous studies, Refs.~\cite{Grin:2007yg, Carenza:2021ebx}, we extrapolate to larger masses, by assuming that the 95$\%$ contour converges to a line determined by the axion free-streaming set by $\lambda_{\rm min}=2\pi/k_{\rm max} \simeq 35$~Mpc, this is shown by the dashed-dotted line in Fig.~\ref{fig:bounds_22}. 

To gain deeper insights into the data presented in Ref.~\cite{Xu:2021rwg} and their implications for light relics, in Fig.~\ref{fig:bounds_22} we have added two lines representing the limits imposed by $\Delta N_{\rm eff}<0.3$ and the hot/warm relic abundance $\Omega_{h}h^2\lesssim 0.002$, labeled `$\Delta N_{\rm eff}=0.3$' and `$\Omega=\Omega_{h}$', respectively. 

These limits are commonly referenced when analyzing Planck 2018 temperature and polarization data. These two constraints are similar to the allowed region in Fig.~\ref{fig:bounds_22}, however, they are weaker for masses around a few eV as well as below $1$~eV. 

Additionally, we have introduced the black-dotted and green-dashed lines illustrating the axion temperature today relative to its mass, considering interactions with gluons and photons, respectively, under the assumption of standard cosmological history. Masses ranging from $0.2$~eV to $18$~eV are ruled out for the gluon interaction, while for the Primakoff interaction, the excluded range narrows to $2.5$~eV to $10.5$~eV.

In particular, the integration of robust LSS data enables more precise constraints, particularly in the low-mass parameter space below the eV scale. Furthermore, it improves our understanding of the mass region between 2~eV and 10~eV, where the data exhibit a pronounced slope, and is much better than the naive estimate of the hot relic abundance.

\section{Thermal axions in nonstandard cosmologies} \label{sec:NSCThermalaxions}

\subsection{Nonstandard cosmologies}
The cosmological history of our Universe prior to BBN is so far totally unknown.
In the so-called standard cosmological scenario, it is assumed that between the end of inflationary reheating and the onset of BBN, the energy density of the Universe was dominated by SM radiation. Additionally, the end of reheating is also taken at a very high scale, well above the typical scales of the processes studied.
However, the energy density of the post-inflationary Universe could have been dominated by something other than SM radiation, resulting in a period of expansion that deviates from standard cosmology~\cite{Allahverdi:2020bys}. The impact on axion DM production during this period has been extensively studied~\cite{Steinhardt:1983ia, Lazarides:1990xp, Kawasaki:1995vt, Giudice:2000ex, Grin:2007yg, Visinelli:2009kt, Nelson:2018via, Visinelli:2018wza, Ramberg:2019dgi, Blinov:2019jqc, Carenza:2021ebx, Bernal:2021yyb, Venegas:2021wwm, Arias:2021rer, Bernal:2021bbv, Arias:2022qjt, Choi:2022btl, Mazde:2022sdx}.
Here, we aim to take advantage of the progress made to obtain smooth decay rates for the coupling of axions to gluons and photons and to re-examine the status of EMD, LTR, kination, and kination-like scenarios. 

\subsubsection*{Early matter domination}
In this setup, the evolution of the background is governed by the system of Boltzmann equations
\begin{align}
    \frac{d\rp}{dt} + 3\, H\, \rp = - \Gphi\, \rp\,, \label{eq:boltzmann1}\\
    \frac{d\rR}{dt} + 4\, H\, \rR = + \Gphi\, \rp\,, \label{eq:boltzmann2}
\end{align}
where $\rR$ and $\rp$ denote the SM radiation and the NSC-driving field energy densities, respectively, and $\Gphi$ is the total decay width of $\phi$ into SM radiation. The Hubble expansion rate $H$ is given by
\begin{equation}
    H = \sqrt{\frac{\rp + \rR}{3\, M_{\rm P}^2}}\,. \label{eq:hubble}
\end{equation}
For an EMD, $\phi$ dominates the energy density of the Universe between $\Teq > T > \Tend$, with $\Tend > \Tbbn$ in order to not spoil the successful predictions of BBN. 
The temperature at the end of the NSC is defined as the stage at which the field has mostly decayed away, {\it i.e.}, when $3\, H(\Tend) \simeq \Gphi$, and therefore~\cite{Chung:1998rq, Giudice:2000ex}
\begin{equation} \label{eq:Tend}
    \Tend^2 = \frac{1}{\pi}\, \sqrt{\frac{10}{\gs(\Tend)}}\, M_P\, \Gphi\,.
\end{equation}

The nonstandard expansion has two different phases, an adiabatic one, where the $\phi$ field dominates the expansion without effectively decaying, and then a nonadiabatic one, where entropy is injected into the SM thermal bath due to the decay of the field. We denote the transition temperature between these periods as $\Tc$. The evolution of the Hubble parameter can be conveniently expressed as~\cite{Giudice:2000ex, Arias:2019uol, Arias:2021rer}
\begin{equation}
    H(T) \simeq
    \begin{dcases}
    H_R(T) &\text{ for } T \geq \Teq,\\
    H_R(\Teq) \left(\frac{\gss(T)}{\gss(\Teq)}\right)^{1/2} \left(\frac{T}{\Teq}\right)^{3/2} &\text{ for } \Teq \geq T \geq \Tc,\\
    H_R(\Tend) \left(\frac{T}{\Tend}\right)^4 &\text{ for } \Tc \geq T \geq \Tend,\\
    H_R(T) &\text{ for } \Tend \geq T.
    \end{dcases}
\end{equation}
This allows us to find the relationship between the scale factor and temperature during the different phases, given by
\begin{equation}
    R(T) =
    \begin{dcases}
        \Rc \left(\frac{\gs(\Tc)}{\gs(T)}\right)^{1/3} \frac{\Tc}{T} &\text{for } T \geq \Tc,\\
        \Rend \left(\frac{\gs(\Tend)}{\gs(T)}\right)^{2/3} \left(\frac{\Tend}{T}\right)^{8/3} &\text{for } \Tc \geq T \geq \Tend,\\
        \Rend \left(\frac{\gs(\Tend)}{\gs(T)}\right)^{1/3} \frac{\Tend}{T} &\text{for } \Tend \geq T,
    \end{dcases}
\end{equation}
where $\Rc \equiv R(\Tc)$ and $\Rend \equiv R(\Tend)$.
The total entropy injected into SM radiation can be estimated to be
\begin{equation}
    \frac{S(\Tend)}{S(\Tc)} = \left(\frac{\gs(\Tc)}{\gs(\Tend)}\right)^2 \left(\frac{\gss(\Tend)}{\gss(\Tc)}\right) \left(\frac{\Tc}{\Tend}\right)^5.
    \label{eq:entropy_inj}
\end{equation}
We note that if a process such as axion decoupling occurs during the nonadiabatic phase, only a portion of the total entropy injection is felt, with \(\Tc\) in this expression being replaced by the decoupling temperature.

\begin{figure}[t]
    \centering
    \begin{subfigure}[b]{0.49\textwidth}
    \centering
    \includegraphics[width=1\textwidth]{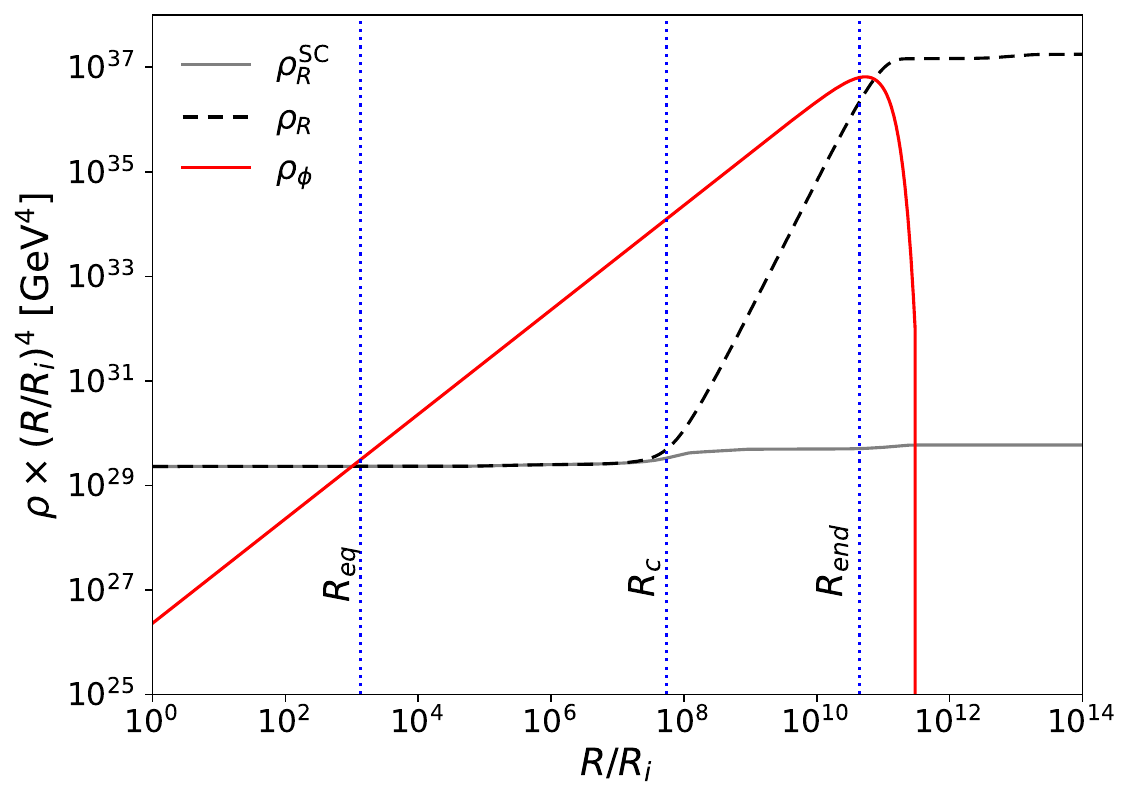}
    \caption{}
      \label{fig:EMD}
    \end{subfigure}
    \hfill
    \begin{subfigure}[b]{0.49\textwidth}
    \centering
    \includegraphics[width=1\textwidth]{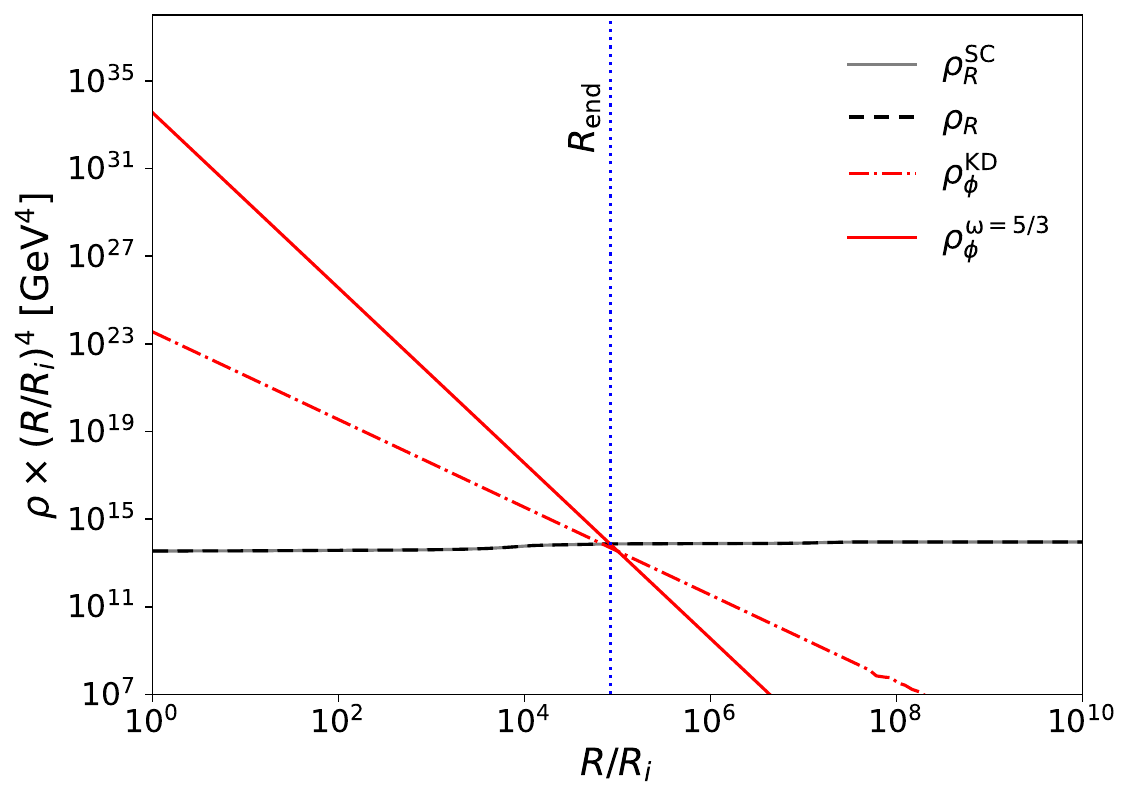}
    \caption{}
      \label{fig:KD}
    \end{subfigure}
    \caption{Evolution of the energy densities for radiation (dashed black) and the $\phi$ field (red) as a function of the scale factor $R$. (a) In EMD with $\Teq = 6.6 \times 10^4$~GeV and $\Tend = 0.02$~GeV. (b) In KD (dot-dashed)  and $\omega=5/3$ (solid) with  $\Tend = 0.02$~GeV, for both cosmologies, the energy density for radiation evolves the same as in SC.}
\end{figure}
The evolution of the energy densities for the SM radiation and $\phi$ as a function of the scale factor is shown in Fig.~\ref{fig:EMD}.
It was obtained numerically by solving Eqs.~\eqref{eq:boltzmann1} and~\eqref{eq:boltzmann2}, for $\Teq = 6.6\times 10 ^4$~GeV and $\Tend = 0.02$~GeV (which implies $\Tc = 0.2$~GeV).
The deviation of the scaling $\rR(R) \propto R^{-4}$ for free radiation between $\Rc$ and $\Rend$ is due to the effective decay of $\phi$ that acts as a source term for SM radiation.

\subsubsection*{Low-temperature reheating}
The low-temperature reheating (LTR) scenario can be recovered from the above in the limit of large $\Tc$, with $\rR(\Tc) = 0$, and identifying $\Tend$ as the inflationary reheating temperature. Therefore, before the onset of the radiation-dominated era, only the nonadiabatic phase is present, corresponding to the inflationary reheating period. In this case, $\phi$ in Eqs.~\eqref{eq:boltzmann1} and~\eqref{eq:boltzmann2} is identified as the inflaton.

\subsubsection*{Kination-like scenarios}
In this case, we assume that the state responsible for the NSC has an energy density that redshifts faster than radiation, having an equation-of-state parameter $\omega > 1/3$. As its energy density eventually becomes subdominant with respect to the SM radiation, it does not have to decay.
The Boltzmann equations describing the system can be written as
\begin{align}
    \frac{d\rp}{dt} + 3\, (1 + \omega)\, H\, \rp = 0\,,\\
    \frac{d\rR}{dt} + 4\, H\, \rR = 0\,.
\end{align}
As $\phi$ does not decay into SM particles, the SM entropy is conserved, and therefore the bath temperature scales as
\begin{equation}
    T(R) = \Tend \left(\frac{\gss(\Tend)}{\gss(T)}\right)^{1/3} \frac{\Rend}{R}\,,
\end{equation}
where $\Rend \equiv R(\Tend)$ corresponds to the scale factor at the moment of the equality $\rR(\Rend) = \rp(\Rend)$.
In this scenario, the Hubble expansion rate can be approximated by
\begin{equation}
    H(T) \simeq
    \begin{dcases}
        H_R(\Tend) \left(\frac{\gss(T)}{\gss(\Tend)} \left(\frac{T}{\Tend}\right)^3\right)^\frac{1+\omega}{2} &\text{ for } T \geq \Tend\,,\\
        H_R(T) &\text{ for } \Tend \geq T\,.
    \end{dcases}
\end{equation}
A typical example of this scenario corresponds to kination~\cite{Spokoiny:1993kt, Ferreira:1997hj}, where $\omega = 1$.
However, larger equation-of-state parameters are also possible. In cosmologies with $\omega = 5/3$, the energy density of the state generating the NSC scales like $R^{-8}$, and therefore the Hubble expansion rate in the NSC phase scales as $H(R) \propto R^{-4}$~\cite{DEramo:2017gpl}.
This equation of state appears in the context of ekpyrotic~\cite{Khoury:2001wf, Khoury:2003rt} or cyclic scenarios~\cite{Gasperini:2002bn, Erickson:2003zm, Barrow:2010rx, Ijjas:2019pyf}; see also Ref.~\cite{Scherrer:2022nnz}.
Fig.~\ref{fig:KD} shows the energy densities for $\phi$ and the SM radiation energy density as a function of the scale factor for $\omega = 1$ (KD) and $\omega = 5/3$, for $\Tend = 0.02$~GeV.

While the analytical estimates provided above offer valuable insights into the role of the NSC in the Universe's expansion, we emphasize that all subsequent calculations have been done by numerically solving the corresponding Boltzmann equations \eqref{eq:boltzmann1} and \eqref{eq:boltzmann2}, and using the Hubble parameter of Eq.~\eqref{eq:hubble}.

\subsection{The population of thermal axions}
\begin{figure}[t]
    \centering
    \begin{subfigure}[b]{0.49\textwidth}
    \centering
    \includegraphics[width=1\textwidth]{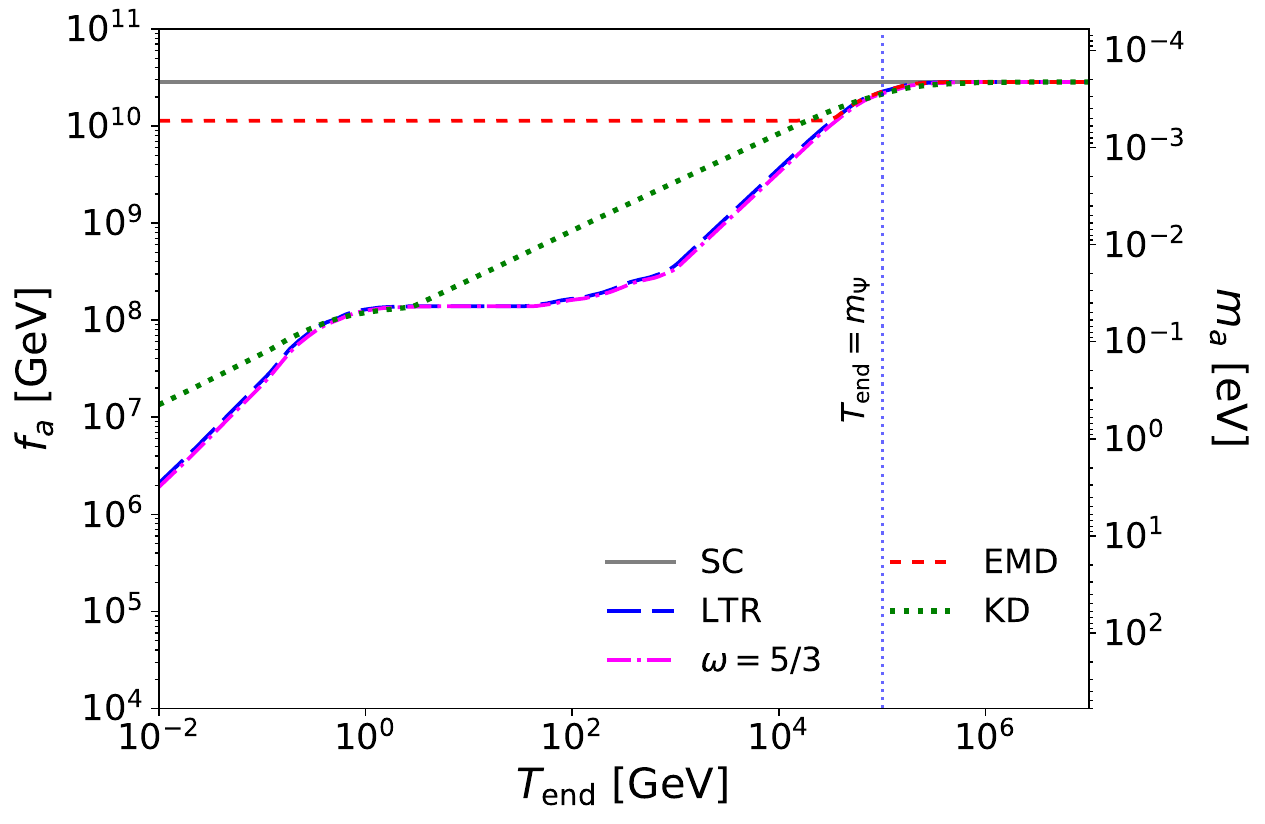}
    \caption{}
    \label{fig:fa_equilibrium}
    \end{subfigure}
    \hfill
    \begin{subfigure}[b]{0.49\textwidth}
    \centering
    \includegraphics[width=1\textwidth]{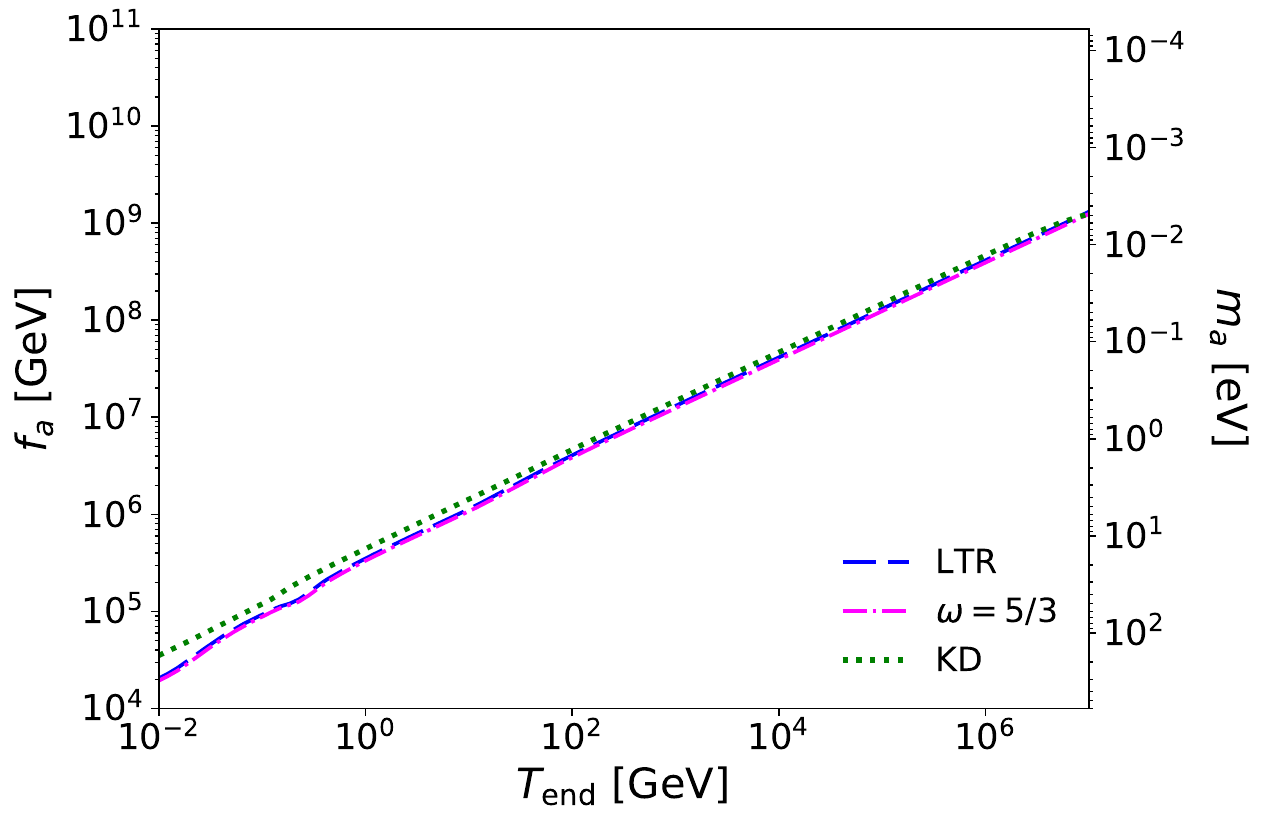}
    \caption{}
    \label{fig:fa_equilibrium_photon}
    \end{subfigure}
     \caption{Parameter space (below the lines) where axions reach chemical equilibrium with the SM plasma, for the SC, EMD (with $\Teq = 10^7$~GeV), LTR, KD and $\omega = 5/3$, and couplings to gluons (left) or photons (right). In the right panel, the SC, EMD, and LTR cases all coincide as described in the text, and therefore we only show LTR to represent these three, while KD and $\omega = 5/3$ are shown separately.
    }
    \label{fig:thermalization}
\end{figure}

In the standard radiation-dominated period, both axion interactions under study (those due to gluon and photon couplings) can be strong enough to produce an axion population in thermal equilibrium. In Fig.~\ref{fig:thermalization} we explore this possibility for a set of non-standard cosmological histories including LTR, EMD, KD, and, for the sake of completeness, a cosmology with $\omega=5/3$. For parameter values in the regions below each curve, thermal equilibrium is achieved at some temperature $T \geq \Tend$ when $\Tend < m_\Psi$, or at some temperature $T \geq m_\Psi$ when $\Tend > m_\Psi$.
In the left panel, corresponding to the axion coupling to gluons, we can see from the solid black line that, for a standard radiation-dominated Universe, thermal equilibrium is guaranteed for $f_a< 3\times 10^{10}$~GeV when the mass of the heavy colored PQ fermion is $m_\Psi=10^5$~GeV.

For the gluon interaction, assuming the validity of the interpolated rate presented in Section~\ref{sec:CouplingGluon}, it is important to note that the highest \(f_a\) at which equilibrium can be achieved is determined by the mass of the PQ fermion, which we have fixed to $m_\Psi=10^5$~GeV.
For temperatures higher than this mass, the rate scales approximately to $T$. Consequently, for $\Tend > m_{\Psi}$, all curves in the left panel converge, since thermalization is achieved in the standard RD period that follows each particular history. For lower ending temperatures, $\Tend\lesssim m_{\Psi}$, the curves depart from standard radiation-dominated cosmology. In the LTR case, with \(H \propto T^4\), the features of the interaction rate at the QCDPT and the heavy fermion mass threshold are clearly seen as the two ``bumps''at the corresponding temperatures, indicating which part of the rate is responsible for ensuring equilibrium. The change in slope at \(\Tend \simeq 1\)~TeV corresponds to a case where equilibrium is instead being achieved by the gluon scattering rate at \(T \simeq \Tend < m_{\Psi}\), rather than the rate near the QCDPT or the heavy fermion mass during the nonadiabatic phase. The case of KD, which has a milder dependence of \(H \propto T^3\) also displays these features, although to a lesser degree, whereas the \(\omega = 5/3\) case, which also has \(H \propto T^4\) essentially mimics LTR. 
Finally, the case of EMD is less clear, due to the prior period of RD before the onset of EMD. This results in an additional dependence on the temperature at the beginning of the EMD, which was taken to be $10^7$~GeV in the figure. For this starting temperature, the EMD curve follows along the LTR case at high \(\Tend\), corresponding to \(H \sim \Gamma\) occurring in the nonadiabatic phase with \(H \propto T^4\), before becoming constant for \(\Tend < 3\times 10^4\)~GeV in which equilibrium is established in the adiabatic phase, with \(H \propto T^{3/2}\). If the start of the EMD is taken to be at temperatures larger than \(10^7\)~GeV, the curve would continue to follow the LTR case down to smaller \(f_a\), while if the start temperature is smaller, the EMD curve approaches the RD line, fully replicating it for \(\Teq < m_\Psi\). 

Based on the \(\Gamma \propto T^3/f_a^2\) scaling of the gluon interaction rate for \(T \lesssim m_\Psi\), and our \(f_a < 3\times 10^{10}\)~GeV value for \(m_\Psi = 10^5\)~GeV, we can estimate the behavior of the curves for different values of the PQ fermion mass. The limiting value of \(f_a\) below which equilibrium is achieved for a standard RD history is approximately given by \(f_a \approx 3\times 10^{10}\,{\rm GeV} \sqrt{m_\Psi / (10^5\,{\rm GeV})}\), which increases with \(m_\Psi\). Therefore, we expect that the upper right portions of the curves will shift upward and to the right as \(m_\Psi\) increases, maintaining their relative positions, while the lower left portions remain unchanged. 
In principle, if \(m_\Psi\) is larger than all inflationary reheat temperatures being considered, one could integrate out the heavy fermions so that they decouple from the low-energy theory. This would result in a continuation of the \(T^3\) scaling of the interaction rate from gluon scattering (as seen in Fig.~\ref{fig:smooth_rate}) to temperatures larger than \(\Tend\) for all \(\Tend\) considered. The feature at \(\Tend \sim m_\Psi\), where all curves converge, would then be replaced by a scaling similar to that seen in the photon interaction discussed below, where larger \(f_a\) can establish equilibrium as long as \(\Tend\) is sufficiently large. However, we find it informative to study the case where \(m_\Psi\) is accessible precisely because it introduces an upper limit on the scale \(f_a\) above which thermalization from the gluon interaction is no longer guaranteed for any cosmological history, independent of the value of \(\Tend\) as long as it is larger than \(m_\Psi\).

For the Primakoff interaction, shown in the right panel of Fig.~\ref{fig:thermalization}, the interaction rate maintains the scaling \(\Gamma_Q \propto T^3/f_a^2\) throughout the cosmological history. For the purposes of determining the largest \(f_a\) that can accommodate equilibrium in a standard RD history, this introduces a dependence on the highest temperature reached in the RD period, namely the inflationary reheat temperature. 
Therefore, the standard RD case is essentially the same as LTR, albeit with a higher reheat temperature. The largest \(f_a\) that allows for equilibration can then be estimated by evaluating \(H_R = \Gamma_Q\) at the inflationary reheat temperature using Eqs.~\eqref{eq:GammaPrimakoff} and~\eqref{eq:H(T)rad}, giving \(f_a \simeq 1.5\times 10^9~{\rm GeV} \sqrt{\Tend / (10^7\,{\rm GeV})}\). In the figure, we show a single curve for both the RD and LTR cases, with \(\Tend\) corresponding to the inflationary reheat temperature. In the case of EMD, where the matter period is preceded by a radiation phase, the curve is the same as in the case of RD/LTR, with the understanding that \(\Tend\) is again the inflationary reheat temperature rather than the end of EMD. The cases of KD and \(\omega = 5/3\) also result in essentially the same curve, but in this case both the Hubble rate and the photon interaction rate scale as \(T^3\) for KD, while \(H \propto T^4\) for \(\omega = 5/3\). Instead of the reheat temperature, \(\Tend\) now indicates the end of the kination-like period, which still corresponds to the maximum temperature reached in the radiation phase. 

We should note that this discussion applies only to finding the maximum value of \(f_a\) below which equilibrium is guaranteed. Here, we have not treated the differences between the histories when it comes to the decoupling temperature. 

\subsection{Signatures of thermal axions formed during NSCs}
In this section, we discuss the changes in observables related to axion thermal production under two different cosmological scenarios: LTR and EMD. We focus on these cosmologies, on the one hand, to make contact with previous works~\cite{Grin:2007yg, Carenza:2021ebx}. On the other hand, cosmological periods with $\omega<0$ will differentiate from EMD mainly in the total amount of entropy injected into the primordial bath. That is, they further dilute the thermal population. Kination and similar cosmologies ($\omega>1/3$) do not deposit extra entropy, as they dilute faster than radiation. Therefore, the only change from SC is a higher decoupling temperature, reaching almost the same constraints as in the standard scenario.

We will constrain the LTR and EMD periods using the data in Fig.~\ref{fig:bounds_22}. When the axion population undergoes thermalization and decouples during an NSC, the decoupling temperature is higher than that of the RD scenario. This is due to the increased Hubble parameter, resulting in a higher dilution caused by the change in the relativistic degrees of freedom in the SM. Fig.~\ref{fig:decouplingT} illustrates the comparison of decoupling temperatures between an RD Universe (solid gray) and different EMD histories for the gluon coupling (left) and photon coupling (right).

However, in the case of EMD and LTR there is also an entropy dilution from the new field $\phi$ when it begins to decay. To incorporate this effect into the observables discussed in Section~\ref{sec:Thermalaxions}, we focus on the axion temperature today, since we can express all observables in terms of this variable; see Eqs.~\eqref{eq:axion_relic} and~\eqref{eq:massless_Neff}. 

We assume that the decoupling occurs at bath temperature $\Tdnsc > \Tend$.
After decoupling, the axion temperature will then continue to redshift as
\be
    T_a = \Tdnsc\, \frac{R_{\rm d, nsc}}{R}.
\ee
To relate this temperature to the current temperature, we consider the nonadiabatic period between $\Tend < T < \Tc$ using the injection of entropy in Eq.~\eqref{eq:entropy_inj}. The axion temperature today is given by
\begin{equation}
    T_{a,0}= T_{0} \left(\frac{\gss(T_0)}{\gss(\Tdnsc)}\right)^{1/3} \times
    \begin{dcases}
        \left[\frac{S(\Tc)}{S(\Tend)}\right]^{1/3} & \text{ for } \Tdnsc > \Tc\,,\\
        \left[\frac{S(\Tdnsc)}{S(\Tend)}\right]^{1/3} & \text{ for } \Tend < \Tdnsc < \Tc\,.
    \end{dcases}
    \label{eq:T0_nsc}
\end{equation}
The second case in the above equation also applies to an LTR cosmology, where $\Tc$ is a very high scale. For decoupling temperatures smaller than $\Tend$, the expression coincides with that of RD, Eq.~\eqref{eq:Td_sc}.

\subsubsection*{Observables in NSC}
We first write down the abundance of axions today, in the case where the decoupling happens during a NSC. We go back to Eq.~\eqref{eq:axion_relic} and replace today's temperature for the one in an NSC scenario, Eq.~\eqref{eq:T0_nsc}. Therefore,
\begin{equation}
    \Omega_a\, h^2
    \simeq 0.02 \left(\frac{m_a}{\rm{eV}}\right) \left(\frac{\gss(T_0)}{\gss(\Tdnsc)}\right) \times
    \begin{dcases}
        \left[\frac{S(\Tc)}{S(\Tend)}\right] &\text{ for } \Tc<\Tdnsc<\Teq\\
        \left[\frac{S(\Tdnsc)}{S(\Tend)}\right] &\text{ for } \Tend<\Tdnsc<\Tc\\
        1 &\text{ for } \Tdnsc<\Tend\,,
    \end{dcases}
\end{equation}
where in the last case, it is implied $\Tdnsc= T_{\rm d}$. As a result of the smaller energy, the bounds relax with respect to the standard scenario.

\begin{figure}[t!]
    \centering
    \includegraphics[scale=0.5]{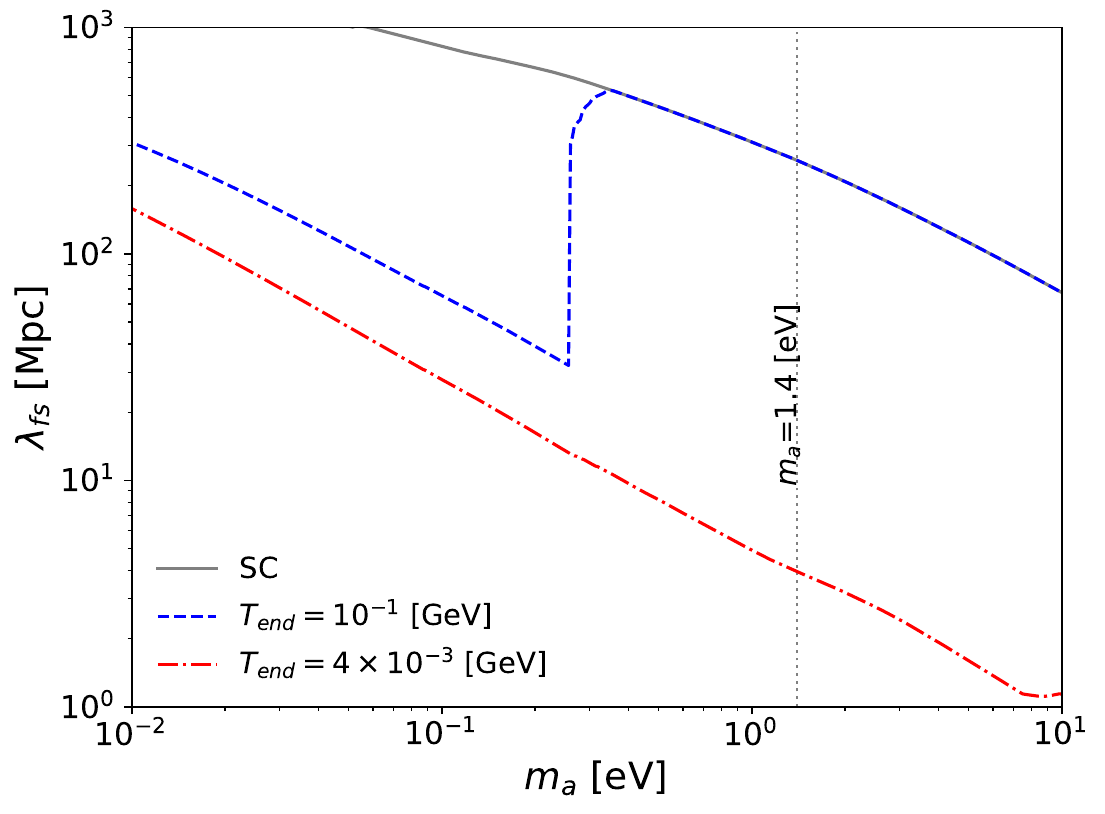}%
    \caption{Comoving free-streaming length evaluated at matter-radiation equality as a function of the axion mass. The continuous gray line corresponds to the standard cosmological scenario. The blue and red lines correspond to EMD cosmologies with $\Teq \simeq 7\times 10^3 $ GeV and $\Tend=0.1$~GeV, $\Tend=0.004$~GeV respectively.  }
    \label{fig:fs_NSC}
\end{figure}

The same strategy can be used for the contribution to the relativistic degrees of freedom during the CMB decoupling and the velocity of thermal axions after they become non-relativistic. 
In the first case, we obtain
\be
    \DNeff \simeq 0.55 \left(\frac{\gss(T_0)}{\gss(\Tdnsc)}\right)^{4/3} \times
    \begin{dcases}
        \left[\frac{S(\Tc)}{S(\Tend)}\right]^{4/3} & \text{ for } \Tc<\Tdnsc<\Teq\,,\\
        \left[\frac{S(\Tdnsc)}{S(\Tend)}\right]^{4/3} & \text{ for } \Tend<\Tdnsc<\Tc\,,\\
        1 & \text{ for } \Tdnsc<\Tend\,,
    \end{dcases}
\ee
while for the velocity, we use Eq.~\eqref{eq:velocity},
\begin{align} \label{eq:vel_nsc}
   \langle v_a\rangle &\simeq 95\, \mbox{km\, s}^{-1} \left(\frac{\mbox{eV}}{m_a}\right) \left(\frac{\gss(T_0)}{\gss(\Tdnsc)}\right)^{1/3} (1+z) \times
    \begin{dcases}
        \left[\frac{S(\Tc)}{S(\Tend)}\right]^{1/3} & \text{ for } \Tc<\Tdnsc<\Teq\,,\\
        \left[\frac{S(\Tdnsc)}{S(\Tend)}\right]^{1/3} & \text{ for } \Tend<\Tdnsc<\Tc\,,\\
        1 & \text{ for } \Tdnsc<\Tend\,.
    \end{dcases}
\end{align}
It is interesting to note that the thermal velocity is the observable that is least affected by the dilution. Therefore, we expect that even though the constraints on axions of small mass (dark radiation) can be severely lifted in non-standard cosmological histories, they can still leave a distinctive imprint through their impact on LSS observations. In effect, we can easily find the redshift at the approximate moment when the decoupled axion becomes non-relativistic -- the analog of Eq.~\eqref{eq:nr_st} -- because it can also be written in terms of today's temperature
\be
    z_{\rm{ nr}} + 1 \simeq \frac{m_a}{2.7 \,  T_{a,0}} \simeq 3436\frac{m_a}{1.4\, \mbox{eV}} \left(\frac{\gss(\Tdnsc)}{14.5}\right)^\frac13 \times 
    \begin{dcases}
        \left[\frac{S(\Tend)}{S(\Tc)}\right]^{1/3} & \Tc<\Tdnsc<\Teq\,,\\
        \left[\frac{ S(\Tend)}{S(\Tdnsc)}\right]^{1/3} & \Tend<\Tdnsc<\Tc\,.
    \end{dcases}
\ee
Note that in the above expression, the entropy injection appears, which is always $\gtrsim 1$. Therefore, the moment the axions become non-relativistic happens earlier if there is a nonadiabatic phase, leading to an earlier transition into a non-relativistic state. This trend has also been observed in Ref.~\cite{Carenza:2021ebx}. As an example, in Fig.~\ref{fig:fs_NSC} we show the free-streaming length at the moment of matter-radiation equality as a function of the axion mass for the gluon coupling. We can see that small axion masses, which decouple very early, have a much larger entropy dilution and thus do not follow the standard RD cosmology (solid gray line) because they became nonrelativistic prior to matter-radiation equality. For $\Tend=100$~MeV (dotted blue) this behavior occurs for masses up to $m_a \lesssim 0.2$~eV and higher masses already decouple after the end of the NSC. For cosmologies with a higher entropy dilution, such as the dash-dotted red line, masses below 10~eV are already non-relativistic at the equality, with a much reduced free-streaming length. Therefore, the axions become CDM-like for observational purposes. For several NSC histories, the thermal axion population can have more warm DM features than hot DM. Therefore, it is expected that they impact the matter power spectrum at high wave numbers $k_{\rm fs}\equiv 2\pi/\lambda_{\rm fs}$. As a consequence, most of their effects appear at the nonlinear scale, where the perturbative analysis breaks down. This distinctive feature can be the key to distinguishing hot relics formed during an RD history from those formed during NSC.

Let us now return to the general picture of the impact of an NSC on the population of relic thermal axions. The energy dilution of the population is the most significant feature of the NSC analyzed here. In the case of LTR, a larger loosening of the constraints will occur as long as the entropy injection is also larger. This situation arises if $\Tdnsc$ is as high as possible compared to $\Tend$. Therefore, we expect the highest relaxation of the bounds for small values of the reheating temperature and small masses. In the case of EMD cosmologies, the decoupling can occur at three different stages of the NSC history: before the decay of the $\phi$ field, during the decay, or afterward. In the first case, the injection of entropy could be the highest possible (and thus loosen the constraints), especially for higher $\Tc$ (or equivalently $\Teq$) and smaller $\Tend$. This will happen for long NSC periods and small masses (high decoupling temperatures). The second possibility has the same outcome as the LTR scenario; therefore, the less constrained scenario comes from small masses and small $\Tend$. The third possibility corresponds to decoupling during an RD scenario, so all constraints from Section~\ref{sec:Thermalaxions} apply.

With all these considerations, we will now explore the parameter space that can be constrained for EMD and LTR cosmologies using the same data of Ref.~\cite{Xu:2021rwg} we used previously for the standard cosmology.

\subsection{Constraints from light massive relics}{\label{sec:constraints}}
\begin{figure}[t]
    \centering
    \includegraphics[scale=0.48]{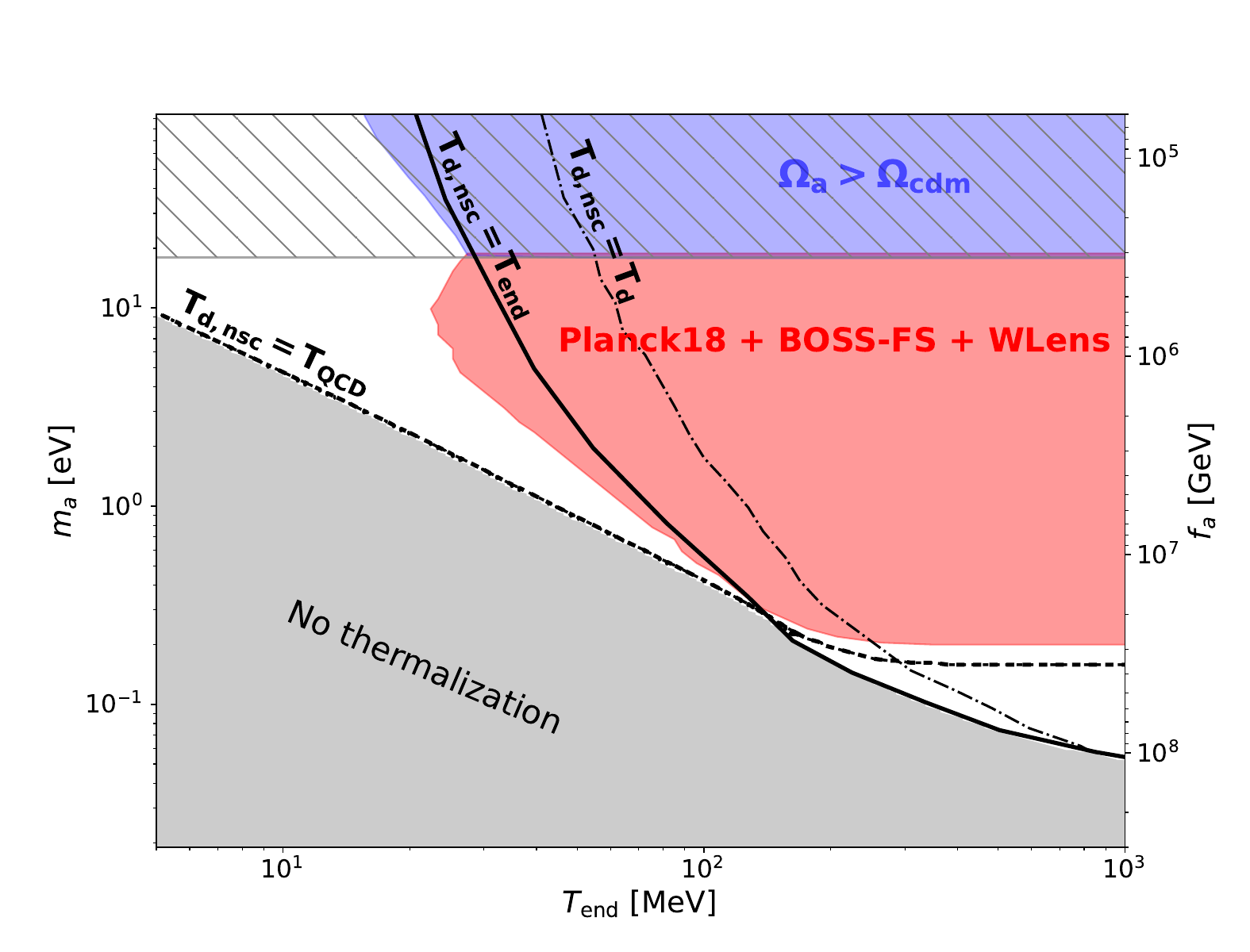}
    \caption{Parameter space as a function of the axion mass/decay constant and the NSC end temperature, $\Tend$, for LTR cosmologies for the axion-gluon coupling. The blue region shows the parameter space where the axion is overproduced with respect to CDM. The red region labeled ``Planck18+BOSS-FS+WLens'', corresponds to the constraints obtained using the CMB, full-shape galaxy and lensing data from Ref.~\cite{Xu:2021rwg}, and is excluded with a 95\% CL. The parameter space where the axions do not achieve full thermal equilibrium is marked in gray and hatched where they decay before today. See the text for more details.}
    \label{fig:bounds_ma_Tend}
\end{figure}
We commence obtaining constraints on the formation of a thermal axion population in the early Universe for LTR cosmologies for the axion-gluon coupling. To do so, we compute for each mass $m_a$ and reheating temperature $\Tend$, today's axion temperature from Eq.~\eqref{eq:T0_nsc} and compare with the data shown in Fig.~\ref{fig:bounds_22}, which we remind the reader includes Planck, full-shape LSS data from BOSS DR12 and weak lensing from CFHTLens. The results appear in Fig.~\ref{fig:bounds_ma_Tend}, where the red area (marked ``Planck18+BOSS-FS+WLens'') is excluded from the data. The gray area (marked ``No thermalization'') is the parameter space where the axions do not fully thermalize, in correspondence with Fig.~\ref{fig:fa_equilibrium}. The blue region corresponds to the parameter space where the abundance of the thermal population exceeds the abundance of CDM today. The streaked region shows the masses above which the axion has a lifetime smaller than the age of the Universe.  As was previously anticipated, the bounds on small axion masses are lifted for cosmologies with a high entropy dilution, resulting in a LTR with small $\Tend$. The condition $\Tdnsc=\Tend$ is shown as a solid black line. Nonetheless, decoupling temperatures slightly smaller than $\Tend$ still happen during the NSC period, as the decay of the new field is not instantaneous. For this, we have also included the black dot-dashed line labeled `$\Tdnsc=\Td$', where the decoupling occurs just at the moment when the radiation content is no longer affected by the NSC. That is, to the right of that line we have decoupling temperatures in SC.

The shape of the regions in the figure can be understood as follows: Higher reheating temperatures correspond to the decoupling during SC; therefore, the bounds are independent of $\Tend$ and are the same as in SC, {\it i.e.} $m_a\lesssim 0.2$~eV.
As the reheating temperature decreases, it eventually reaches the line $\Tend=\Tdnsc$, first at smaller masses that have higher decoupling temperatures. Decreasing $\Tend$ further leads to no constraints for the smallest axion masses constrained in SC, due to the high dilution lowering the axion's temperature below the data's reach for those masses (for smaller $\Tend$, we also see that the small-mass region does not thermalize). For masses around and above an eV, the bounds of the data shown in Fig.~\ref{fig:bounds_22} become stronger, due to the variety of LSS data used. This allows us to constrain reheating temperatures down to 25-30~MeV for that mass range.  
Let us point out that the constrained red-colored region lies very close to the line $T_{\rm{d, nsc}}=\Tqcd$, meaning reaching masses where the decoupling temperature is only just below the QCD temperature $\Tqcd \simeq 150$~MeV.
This reinforces the relevance of using a complete and smooth interaction rate around that temperature.

\begin{figure}[t]
    \centering
    \includegraphics[scale=0.48]{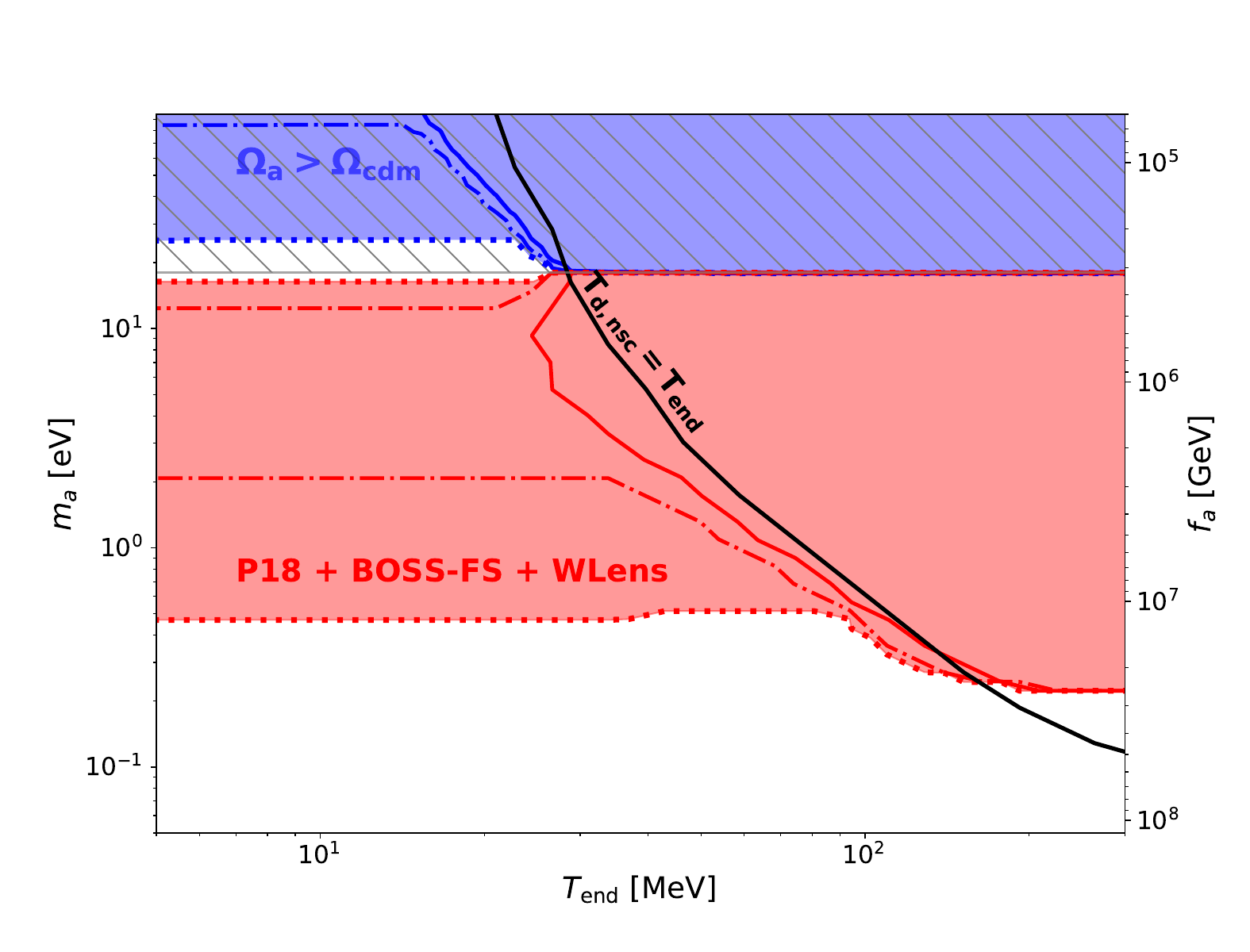}
    \caption{Parameter space as a function of the axion mass/decay constant and the NSC end temperature, $\Tend$, for EMD cosmologies for the axion-gluon coupling. Constrained colored areas have the same meaning as in Fig.~\ref{fig:bounds_ma_Tend}. We show three different scenarios: the regions enclosed by the solid red and blue lines have a fixed ratio of $S(\Tc)/S(\Tend) = 2\times 10^{-4}$, those enclosed by the dot-dashed lines have $S(\Tc)/S(\Tend) = 0.2$, while those enclosed by the dotted lines have $S(\Tc)/S(\Tend) = 0.7$.  
    White regions are allowed.}
    \label{fig:EMD_bounds}
\end{figure}
For EMD cosmologies, we have chosen to show our constraints according to the total amount of dilution possible from the EMD, characterized by $S(\Tc)/S(\Tend)$ given in Eq.~\eqref{eq:entropy_inj}. Therefore, to keep that ratio fixed, we vary $\Tend$ and also $\Tc$ (or equivalently $\Teq$) for each axion mass. In Fig.~\ref{fig:EMD_bounds} we show three different scenarios for the gluon coupling: the red and blue regions bounded by solid lines correspond to EMD cosmologies with a dilution factor of $S(\Tc)/S(\Tend)=2\times 10^{-4}$, while the regions bounded by dot-dashed and dotted lines correspond to $S(\Tc)/S(\Tend)=0.2$ and $S(\Tc)/S(\Tend)=0.7$, respectively. We see that in the first high-diluted scenario, we recover the same results as in LTR Fig.~\ref{fig:bounds_ma_Tend}. This means that the decoupling for all masses that can be probed happens either during the nonadiabatic phase or in SC. Masses that decouple at higher temperatures are not constrained because of the strong dilution. The second and third scenarios, which correspond to shorter EMD cosmologies with less total entropy dilution, can be achieved with $\Tc$ and $\Tend$ being close to each other.\footnote{Another way of having small dilution without requiring $\Tend\sim \Tc$ is with EMD periods where the new field does not fully dominate over radiation. We do not consider such cases here.} The shape of the parameter space in these scenarios in Fig.~\ref{fig:EMD_bounds} can be understood as follows. For high $\Tend$, where all three cases converge, the bounds from SC are recovered. As the temperature decreases, it eventually reaches $\Tend \sim \Tdnsc$, and the dilution begins to become important. As $\Tend$ keeps decreasing and $\Tdnsc<\Tc$, the dilution factor is given by $S(\Tdnsc)/S(\Tend)$, which is smaller than the total dilution factor $S(\Tc)/S(\Tend)$, so the constraints are stronger. This can be seen for $\Tend\gtrsim 35-40$~MeV in the second case. For smaller $\Tend$ the decoupling -- especially for smaller masses -- occurs during the adiabatic phase of the EMD. In that scenario, the dilution factor is the highest possible for all $\Tend$, and the bound flattens.

\begin{figure}[t]
    \centering
    \begin{subfigure}[b]{0.49\textwidth}
    \centering
\includegraphics[scale=0.30]
{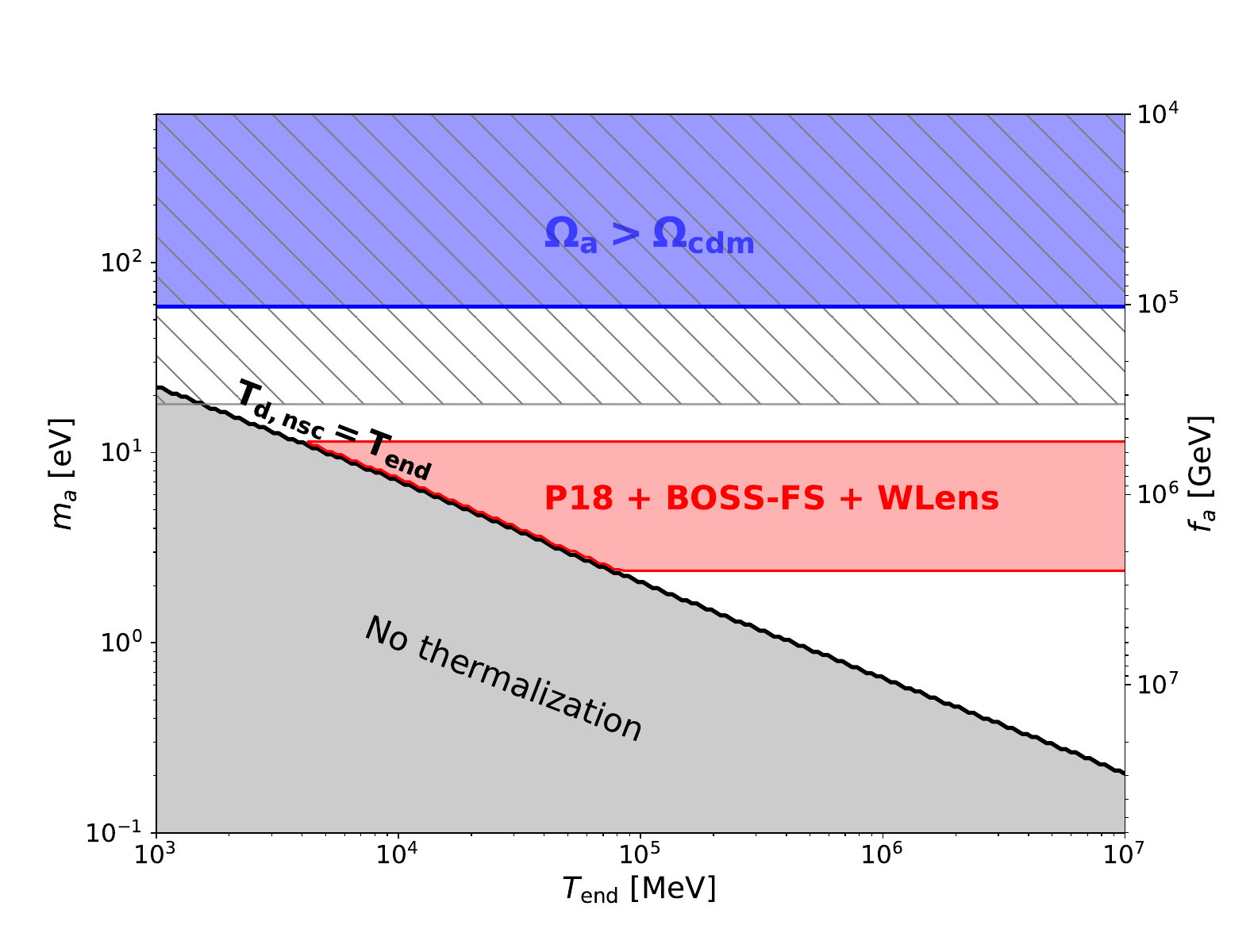}
\caption{}
\end{subfigure}
\hfill
\begin{subfigure}[b]{0.49\textwidth}
\centering
\includegraphics[scale=0.30]
{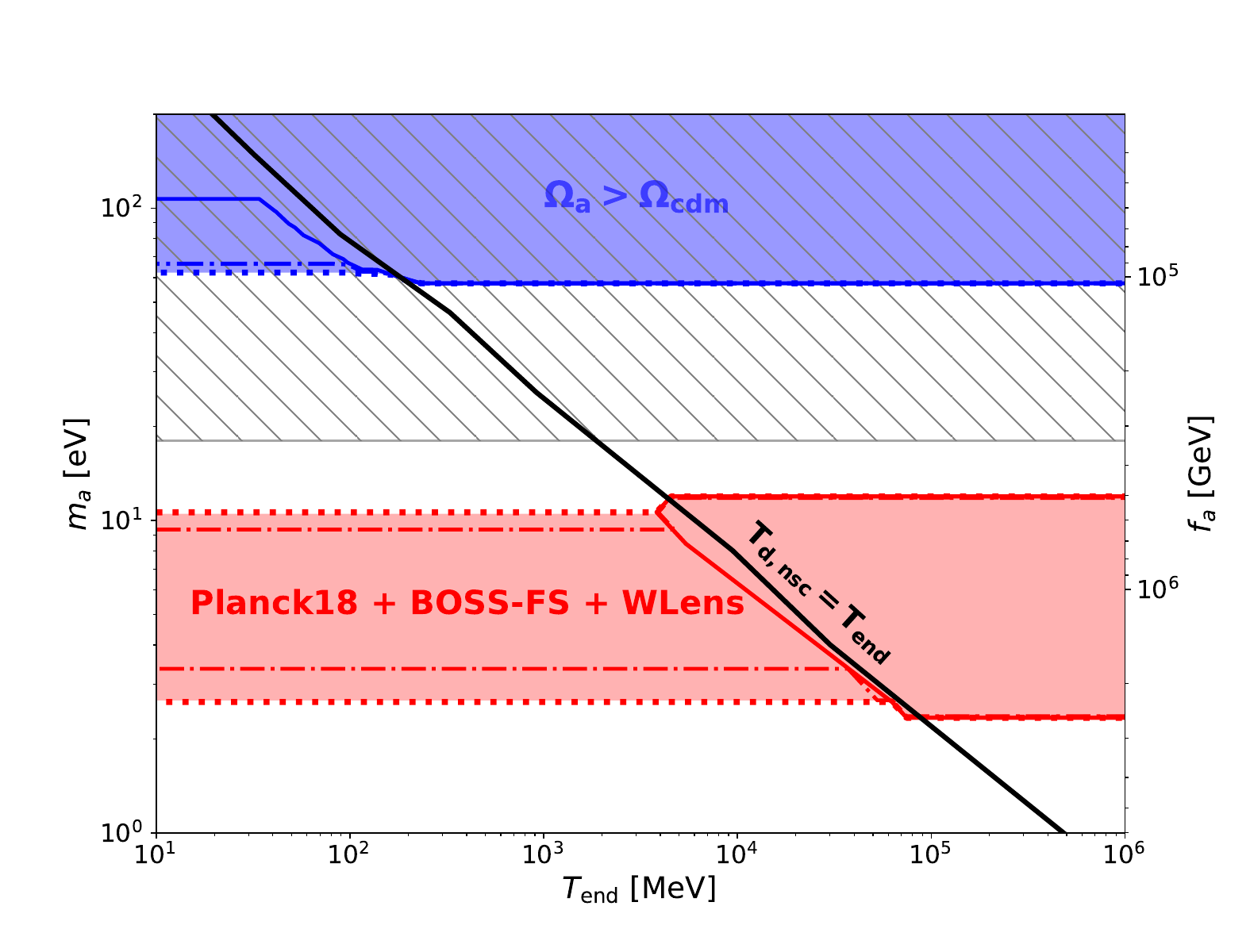}
\caption{}
\end{subfigure}
\caption{Parameter space for the Primakoff interaction in LTR (left) and EMD (right), the colored areas have the same meaning as in Fig.~\ref{fig:bounds_ma_Tend}. (a) The axion never decouples during LTR because of the Hubble parameter's temperature dependence. Once standard cosmology is regained ($T<\Tend$), decoupling proceeds as in the standard case. (b) We compare three different EMD cosmologies: the regions enclosed by the solid red and blue lines have a fixed ratio of $S(\Tc)/S(\Tend)=0.2$, those enclosed by the dot-dashed lines have $S(\Tc)/S(\Tend)=0.4$, while those enclosed by the dotted lines have $S(\Tc)/S(\Tend)=0.7$. White and gray regions are allowed.
}
\label{fig:NSC_photon}
\end{figure}

Finally, in Fig.~\ref{fig:NSC_photon} we show the corresponding plots for the axion-photon coupling. The left panel shows the case of LTR, where it was already anticipated that thermalization will be highly delayed for this interaction. Therefore, for high reheating temperatures, the bounds are the same as those of SC, as the decoupling happens during that period. As the reheating temperature decreases, it reaches the $\Tdnsc=\Tend$ line (which occurs at temperatures much higher than the gluon counterpart). Smaller reheating temperatures do not allow for axion thermalization, so they cannot be constrained by our analysis. In the right panel, we again show EMD cosmologies with a fixed total dilution factor. The photon coupling has much higher decoupling temperatures than the gluon coupling. For that reason, the smallest dilution taken is $S(\Tc)/S(\Tend)=0.2$, enclosed by solid red and blue lines. Smaller dilution factors lead to decoupling in SC and those bounds apply. In this case, the result is not identical to the LTR case due to the difference in thermalization. For this scenario, we can constrain a slim parameter space where the decoupling occurs during the EMD era, and it is better for masses around the eV scale, thanks to the data used.   
By reducing the dilution, we obtain the dot-dashed and dotted lines, with total entropy dilution factors of 0.4 and 0.7, respectively. The same feature as in the case of the gluon coupling is obtained, in the sense that for $\Tend$ below $\Tdnsc$, the dilution factor is given by $S(\Tdnsc)/S(\Tend)$, which is smaller than the total factor, leading to a stronger constraint. As $\Tend$ continues to decrease, the decoupling occurs in the adiabatic phase, and the bound becomes independent of $\Tend$.

\section{Coexistence of cold and hot populations}
\label{sec:co-existence}
An intriguing question is whether axions can leave imprints in cosmology from both a cold dark matter (CDM) population and also a ``thermally" induced one, in the sense that it emerges from the primordial bath.

In standard cosmology, the CDM population arises around the QCDPT, with a mass range of $10^{-6}-10^{-4}$ eV~\cite{Dine:1982ah} from the misalignment mechanism. 
However, a complete thermal population of axions can only emerge due to the gluon interaction for masses above $m_a \sim 10^{-4}$~eV, as we have seen in Fig.~\ref{fig:thermalization}, for \(m_\Psi = 10^5\)~GeV. Thus, to have a co-existence of a CDM population from misalignment and a thermal population from the gluon interaction at masses around $\mu$eV, the latter must be produced via freeze-in. But for such small masses the energy density is negligible (see e.g. Fig.~1 of Ref.~\cite{Archidiacono:2015mda}), which makes them unobservable. On the other hand, from the photon coupling an axion mass $m_a\sim 10~\mu$eV decouples from the bath at around $\Td\sim 10^8$~GeV, much earlier than the oscillation temperature of the axion field.\footnote{The CDM condensate is safe because the inverse-Primakoff process $a+e^{\pm}\rightarrow \gamma+e^{\pm}$ is highly suppressed due to the high energy required for the incoming electron to produce a photon. On the other hand, processes like $\gamma+a+e^{\pm}\rightarrow \gamma+e^{\pm}$ are also suppressed because the coupling to two photons contains an axion derivative. Thus, the amplitude of the process is proportional to $m_a$ and probabilities to $m_a^2$. See, for instance, Ref.~\cite{Arias:2012az}.}  

This is not the case in NSC scenarios with $\omega\geq 1$. On the one hand, from Fig.~\ref{fig:thermalization} we can see that the thermalization of the axions can be largely delayed to higher masses by decreasing $\Tend$. On the other hand, the misalignment production of CDM axions is also altered in cosmologies with $\omega>1/3$, shifting the correct relic abundance to higher masses. Those cosmologies do not feature entropy injection into the thermal bath; the only net effect is to delay the oscillation of the axion field, leading to an overproduction of the relic abundance for the classical axion CDM mass window~\cite{Visinelli:2009kt, Arias:2021rer}.  

In this section, we are interested in finding the parameter space where two populations of QCD axions can actually co-exist in sizable amounts: one cold, via the misalignment mechanism, and the other originating from interactions in the bath, from the coupling with gluons. 
We will use the results of Ref.~\cite{Arias:2021rer} where it was found that for cosmologies with $\omega=5/3$, the misalignment mechanism produces the right amount of CDM observed today for masses smaller than or around the eV range for $\Tend \gtrsim 4$~MeV. In principle, the coexistence could also happen for the $\omega=1$ KD cosmology, but since the correct CDM abundance takes place for smaller masses, the abundance (and therefore impact) of hot axions will be fairly small to escape detection. We will come back to this later in the section.
We will only focus on the axion-gluon coupling because, for the photon coupling, thermalization is largely delayed to masses well above the eV for the range of $\Tend$ we are interested in, according to Fig.~\ref{fig:fa_equilibrium_photon}.

In order to find the yield of axions produced through freeze-in, we numerically solve the Boltzmann equation
\begin{equation}
     \frac{d\,n_a}{dt}+3Hn_a= -\Gamma_a \left(n_a-n_a^\text{eq}\right),
\end{equation}
with the change of variables $Y\equiv n_a/s$. Then, the relic density is simply found as $\rho_a=Y_{\infty}\, s_0\,  m_a$, where $Y_{\infty}$ is the yield long after the freeze-in. 

For the axion-gluon coupling, co-existence can occur if axions do not fully thermalize due to their active interactions, i.e. $\Gamma<H$, otherwise this would lead to the disappearance of the condensate. The population built through freeze-in respects the constraints on hot/cold relics that we have analyzed in the previous section. However, it is not possible to use the results of Fig.~\ref{fig:bounds_22}, as they are only valid for thermal relics, which is not the case considered here. 
Nevertheless, in order to give an educated guess on whether the co-existence parameter space could be allowed from cosmology, we use constraints on non-thermal light sterile neutrinos presented in Ref.~\cite{Acero:2008rh}. There, a relic population of sterile neutrinos was assumed to be produced nonthermally through the Dodelson-Widrow mechanism~\cite{Dodelson:1993je}.
There it is used that despite our ignorance about the phase space distribution, the three observables, $\DNeff, \Omega_a h^2$, and $\langle v_{a,0}\rangle$, satisfy a constraint equation given by
\be
    \langle v_{a,0}\rangle = \frac{\int \frac{p}{m_a}\, f(p)\, d^3 p}{\int f(p)\, d^3 p} \simeq 5.618 \times 10^{-6}\, \frac{\DNeff}{\Omega_a h^2}\,.
\ee
Above it has been assumed that the particles are relativistic/semi-relativistic at photon decoupling, which is a fair assumption for masses around an eV.  For a nonrelativistic relic today, which is still distinguishable from the CDM, the velocity should satisfy $\langle v_{a,0}\rangle \gtrsim 1$~km/s. From the Boltzmann equation, we get the relic abundance of today together with the contribution to $\DNeff$. For their analysis, Ref.~\cite{Acero:2008rh} used WMAP5 data plus small-scale CMB, SDSS LRG data, SNIa data from SNLS, and Lyman-$\alpha$ (conservative) from VHS. 

\begin{figure}[t!]
    \centering
    \includegraphics[scale=0.48]{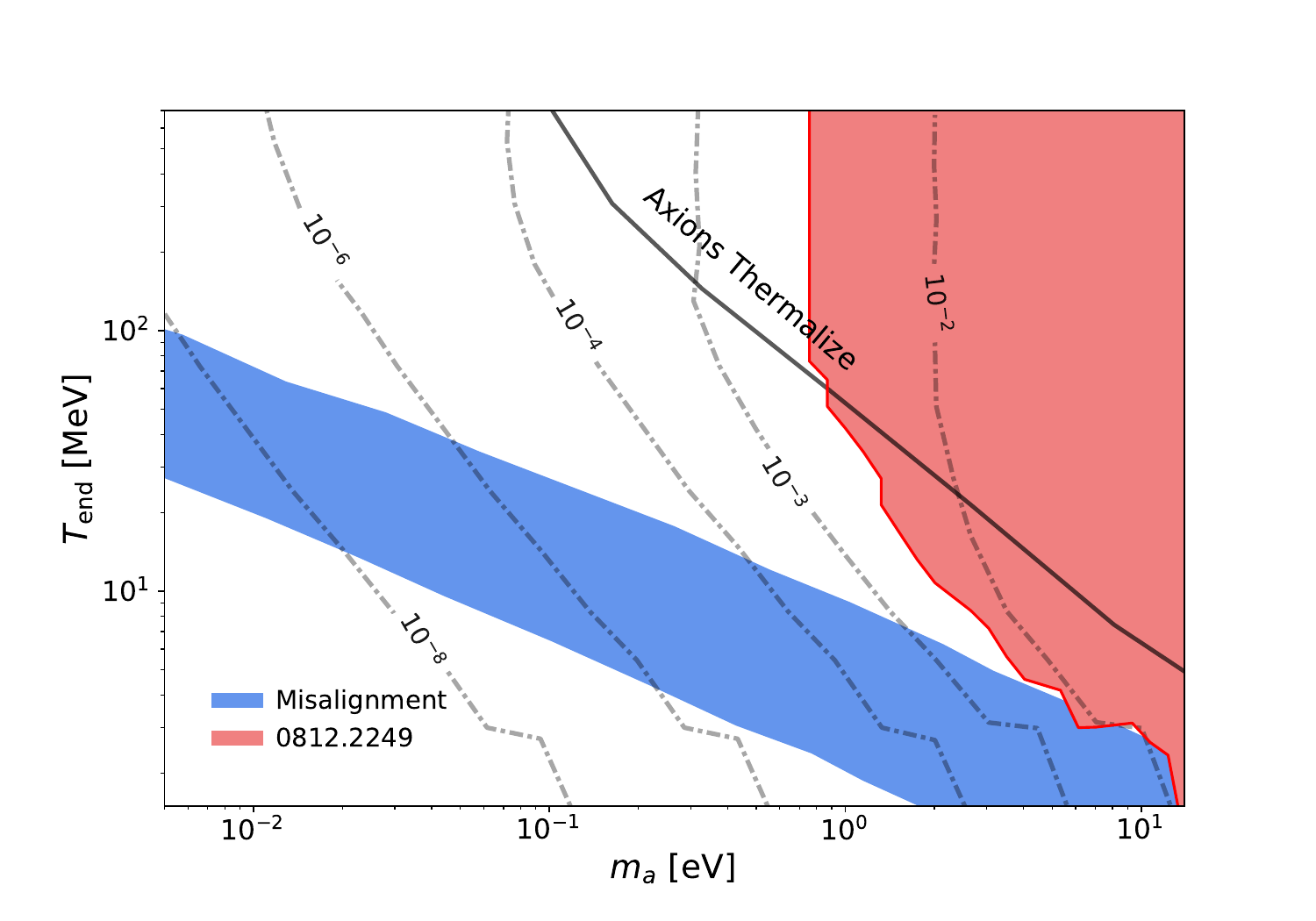}
    \caption{Parameter space of $\Tend$ vs $m_a$ where a misalignment and a thermal (freeze-in) population of QCD axions can co-exist for a cosmology with $\omega=5/3$. The blue region shows the parameter space where the CDM relic abundance can be produced with natural initial misalignment angles (see the text for details). The red region corresponds to the bounds on the sterile neutrinos of Ref.~\cite{Acero:2008rh} applied to the QCD axion. To the right of the solid black line axions produced via the gluon coupling thermalize. The dot-dashed lines show isocontours of relic abundance $\Omega_a h^2$, marked with the corresponding value. White regions are unconstrained.}
\label{fig:co-existence}
\end{figure}
In Fig.~\ref{fig:co-existence} we show the parameter space where the co-existence of a misalignment and a freeze-in population can happen for a non-standard cosmology with $\omega=5/3$ in terms of the axion mass and the temperature where the NSC ends, $\Tend$. The blue band corresponds to the parameter space where axions produced through the misalignment mechanism can account for the entire observed DM today,  assuming a natural range of initial angles, $\theta \in [0.5, 1.8]$, see Ref.~\cite{Arias:2021rer} for details.  
Masses above the solid black line achieve full thermal equilibrium and those below them freeze in. Contours of equal thermal abundances $\Omega_a h^2$ are indicated in dot-dash lines with the corresponding value.

As we can see, most of the parameter space where co-existence happens is unconstrained. However, axions with masses below the eV scale have seemingly small relic abundances, which could lead them to be undetected by current and future surveys. We have checked that all the parameter space scanned in Fig.~\ref{fig:co-existence} corresponds to velocities today higher than $1$~km/s. Therefore, even though these axions contribute to the total CDM abundance, they are still distinguishable from CDM.
The mass range where both populations coexist, with a significant abundance of axions from freeze-in ($\Omega_a h^2\gtrsim 10^{-4}$), is within the sensitivity range of experiments such as IAXO~\cite{IAXO:2019mpb}, BREAD~\cite{BREAD:2021tpx}, and LAMPOST~\cite{Chiles:2021gxk}.

Finally, a comment on the possibility of generating a similar scenario during kination is worth mentioning. From Fig.~\ref{fig:fa_equilibrium}, it can be observed that for the smallest ending temperature allowed by BBN, $\Tend\sim 10$~MeV, masses above $m_a\sim 0.5$~eV thermalize for the axion-gluon coupling in KD. On the contrary, a population of CDM axions will form for that scenario, for $\Tend\sim 10~$MeV, for masses between $10^{-3}-10^{-2}$~eV  (see Fig.~6 from Ref.~\cite{Arias:2021rer}). Hence, it seems possible that the CDM population can co-exist with a nonfully thermal one in KD. However, the relic abundance of the latter is expected to be similar to or slightly higher than the one for a $\omega=5/3$ cosmology.  This is because, despite $\omega=5/3$ having a higher slope than KD, they converge in the vicinity of $\Tend$. This leads to the yield of KD being slightly higher than $\omega=5/3$. From Fig.~\ref{fig:co-existence} we see that the relic abundance is much smaller than $\Omega_a h^2\sim 10^{-8}$ for axion masses $m_a\lesssim 10^{-2}$~eV, for $\Tend\sim$~few MeV. Thus, even though co-existence is possible during KD, the thermal bath population is not abundant enough to be detectable in the near future.

\section{Summary and conclusions}
\label{sec:Conclution}
In this study, we have explored the generation of axions from the thermal bath by their interactions with gluons and photons during cosmological eras characterized by expansion rates different than the standard radiation-dominated scenario, the so-called non-standard cosmologies. 
Our focus was on KSVZ-like models, where axions lack direct couplings to fermions.

We started in Section~\ref{sec:Thermalaxions} with a comprehensive analysis of the formation of a fully thermalized population in standard cosmology and the potential signatures that these particles could leave in cosmological data. Massless or nearly massless axions are well characterized by their contribution to $\DNeff$ and $\sum m_\nu$. On the other hand, light but massive axion relics, while non-relativistic in the present epoch, still maintain non-zero temperatures. This characteristic feature makes it possible to distinguish them from the majority of cold relics, offering a novel approach to identifying tiny relics in cosmological data. Unlike CDM, these thermal relics possess thermal velocities that hinder their clustering beyond a characteristic free-streaming scale. Consequently, they exert a discernible influence on the growth of matter fluctuations, making it possible to detect them through the study of their impact on the large-scale structure of the Universe. Finally, we introduced the data to be used to constrain the fully thermal axion population. We relied on the latest constraints derived from the temperature-mass parameter space for massive light relics once in thermal equilibrium, as reported in Ref.~\cite{Xu:2021rwg}. These constraints were obtained through a comprehensive analysis that combined BOSS DR12 full-shape galaxy data, Planck 2018 temperature polarization and lensing anisotropies, and CFHTLens galaxy-galaxy ellipticity correlations. We applied those restrictions to SC in Fig.~\ref{fig:bounds_22}, to find excellent agreement with previous results reported in the literature. 

In Section~\ref{sec:NSCThermalaxions} we started by making a detailed analysis of the thermalization of axions in standard and nonstandard scenarios for both couplings to gluons and photons. For the gluon coupling we found that the mass of the heavy PQ fermion sets an important upper limit on the maximum value of the scale \(f_a\) that can lead to thermalization in a given history. This is in contrast to the photon coupling which becomes increasingly efficient at thermalization as the maximum temperature of the RD phase increases.

Subsequently, we moved towards analyzing the scenario where axions thermalize in LTR and EMD cosmologies for the gluon and photon couplings. We assessed the influence of the NSC scenarios on three key parameters: the relic abundance, $\Omega_a h^2$, the number of extra relativistic degrees of freedom, $\DNeff$, and the free-streaming length, characterized by the thermal velocity, $\langle v_a\rangle$. Although the impact of light massive relics is inherently complex, in many models with non-thermal distortions, the observable effects can be effectively parameterized using these three quantities with considerable accuracy, as discussed in Ref.~\cite{Cuoco:2005qr}.

Axion thermalization in LTR cosmologies for the gluon coupling has been addressed in the literature before, in Refs.~\cite{Grin:2007yg, Carenza:2021ebx} and our results are in good agreement. The parameter space where the thermal population can exist without restrictions opens up, especially between 0.2~eV $\lesssim m_a\lesssim 10$~eV for cosmologies with reheating temperatures smaller than 100~MeV. Our study expands the previous findings in the following way: we use interaction rates for both gluon and photon couplings that are continuous across the QCDPT. The former allows us to scan smaller axion masses that have higher decoupling temperatures. 
However, the data used to constrain the NSC include the full-shape from BOSS and weak-lensing data as an additional component with respect to Refs.~\cite{Grin:2007yg, Carenza:2021ebx}. Due to the above, in contrast to previous works, we are able to constrain LTR scenarios where the axion freeze-out occurs during the nonstandard expansion, for masses between $1~\mbox{eV}\lesssim m_a \lesssim 15$~eV and for reheating temperatures around 25-30~MeV.We present a deeper insight on the complementarity of our limit with previous ones in Appendix~\ref{sec:appendix}.

A thermal population in EMD cosmologies has not been studied before, to the best of our knowledge. We have established our constraints based on the total amount of entropy injected into the thermal bath. First, for the axion-gluon coupling, we have found that in the case of high dilution, the EMD scenario has the same constraints as LTR. This is because due to the decreased abundance and velocity, the data can only constrain masses that decouple during the nonadiabatic phase. As the total entropy injection decreases, it is possible to probe masses that decouple during the adiabatic phase of the EMD, resulting in smaller $\Tend$ temperatures. Our results allow us to easily extrapolate EMD with other entropy injection factors.

Next, in Section~\ref{sec:NSCThermalaxions}, we repeated the analysis for the axion-photon coupling. Our results show that it is only possible to constrain LTR cosmologies with $\Tend>\Tdnsc$ because otherwise, axions do not thermalize, as seen in Fig.~\ref{fig:fa_equilibrium_photon}. On the other hand, for EMD cosmologies, we have found that, due to the high decoupling temperature (see Fig.~\ref{fig:decouplingT}), highly diluted cosmologies hide the population completely, even at high $\Tend$. Only for factors $S(\Tc)/S(\Tend)\gtrsim 0.2$,  constraints appear. We emphasize that a key feature to search for signatures of NSC is their effect on the free-streaming length. In contrast to the severe effects on the contribution to the effective number of neutrinos and the relic abundance, the velocity of the thermal relic is less affected. This means that they can still play an important role in their impact on the formation of LSS. Such an impact is shifted to smaller masses than in SC because of the earlier transition to non-relativistic states. Upcoming large-scale structure surveys such as the Dark Energy Spectroscopic Instrument (DESI)~\cite{DESI:2016fyo}, the Vera Rubin Observatory~\cite{LSST:2008ijt}, and Euclid~\cite{EUCLID:2011zbd}, together with the next generation of CMB experiments will play a major role in either discovering or placing severe constraints.

Finally, in Section~\ref{sec:co-existence} we explored the possibility of having axion signatures from both cold dark matter and hot/warm populations. Axions are naturally produced as a cold condensate by the misalignment mechanism that, in a SC, is well motivated in the mass range around the $\mu$eV. However, for that mass range, axions do not thermalize for the gluon interactions and decouple much earlier for the photon coupling. However, it has been pointed out in the literature that for cosmologies driven by $\omega>1/3$ the formation of a CDM condensate, with the right characteristics demanded by observations, occurs at higher masses. In particular, for $\omega=5/3$ it can reach the eV mass range. We have found that around that mass, a non-negligible population of hot/warm axions from the thermal bath that is not in conflict with cosmological constraints can also exist. A dedicated analysis could constrain a larger portion of the parameter space than the one we showed, as we used data for nonthermal sterile neutrinos that are outdated. The parameter space in which the two populations can coexist is in the range of laboratory experiments, such as IAXO, BREAD, and LAMPOST.

\section*{Acknowledgments}
We are thankful to Javier Redondo, Joerg Jaeckel, and David J.~E.~Marsh for their valuable comments and feedback. PA thanks P. Toro for her valuable help with our workstation. PA thanks AstroCeNT for their hospitality during the completion of this work. PA and MV acknowledge support from FONDECYT project 1221463 and DICYT 042231AR$\_$Postdoc. This work is supported by the grant ``AstroCeNT: Particle Astrophysics Science and Technology Centre" carried out within the International Research Agendas programme of the Foundation for Polish Science financed by the European Union under the European Regional Development Fund. NB received funding from the Spanish FEDER / MCIU-AEI under the grant FPA2017-84543-P. This article is based upon work from COST Action COSMIC WISPers CA21106, supported by COST (European Cooperation in Science and Technology).

\appendix
\section{Appendix}\label{sec:appendix}

\subsubsection*{Changes in the exclusion limit of the axion-gluon coupling by considering a continuous smooth interaction rate}
\begin{figure}[t!]
    \centering
    \includegraphics[scale=0.48]{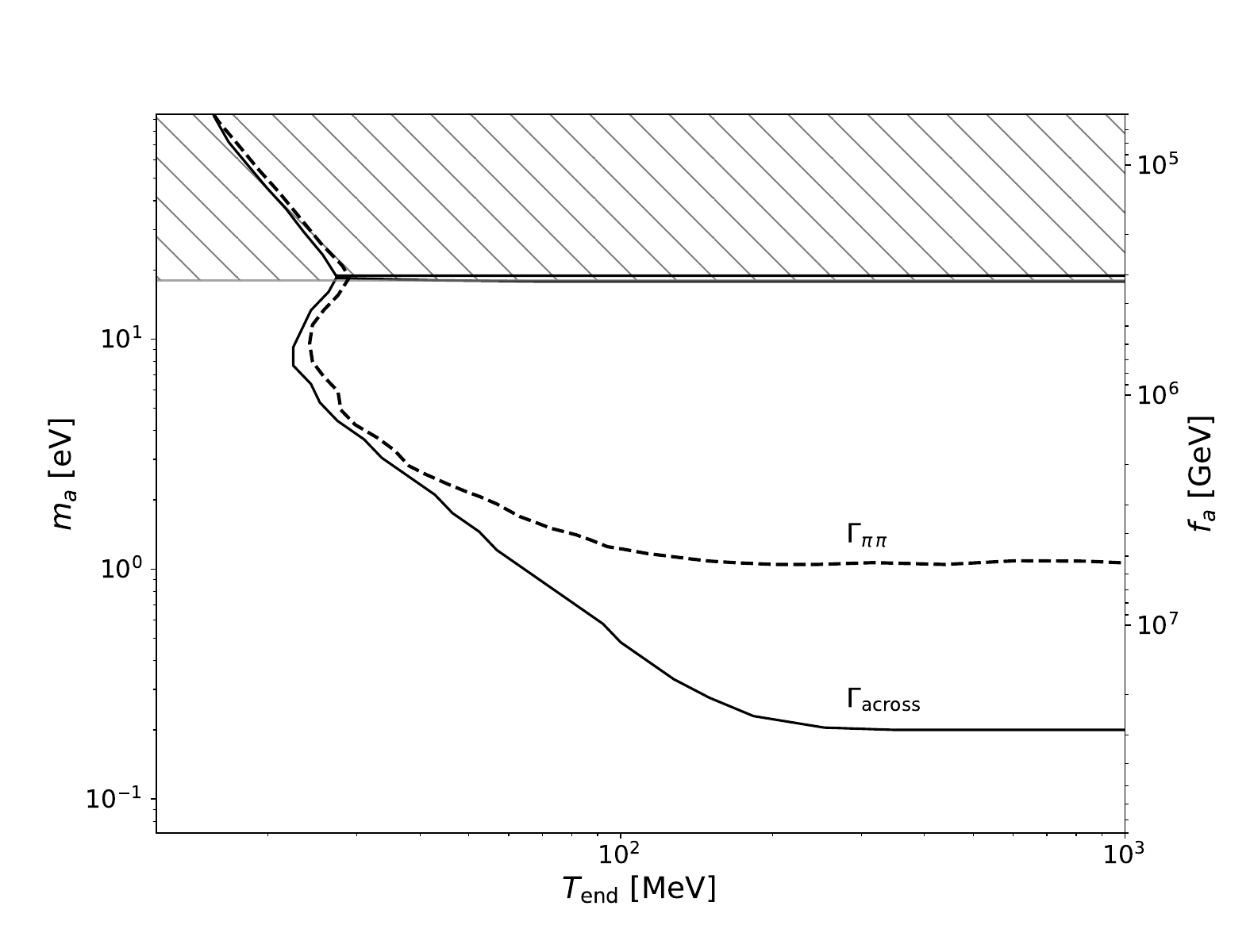}
    \caption{Parameter space as a function of the axion mass/decay constant and the NSC end 
    temperature, $\Tend$, for LTR cosmologies for the axion-gluon coupling. The continuous solid black line is the bound using the smooth interaction rate found in Ref.~\cite{DEramo:2021psx} and is labeled `$\Gamma_{\rm across}$'. Instead, the bound using the axion-pion interaction rate from Eq.~\eqref{eq:rate} is shown as the dashed black line, labeled `$\Gamma_{\pi\,\pi}$'. Both exclusion regions use the LSS data from Ref.~\cite{Xu:2021rwg}.}
\label{fig:rate_comparison}
\end{figure}
In Subsection~\ref{sec:constraints}, we discuss the importance of employing a smooth rate around the QCDPT, as the data are sensitive enough to constrain decoupling temperatures nearing approximately 150~MeV. Here, we aim to refine this assertion by quantitatively comparing the constraint depicted in Fig.~\ref{fig:bounds_ma_Tend}, utilizing the smooth rate derived in Ref.~\cite{DEramo:2021psx,DEramo:2021lgb}, with the one obtained using the rate presented in Eq.~\eqref{eq:rate}, the validity of which has been challenged due to unitarity concerns above $T\sim 62$~MeV~\cite{DiLuzio:2021vjd}. This comparison is shown in Fig.~\ref{fig:rate_comparison}. We observe that the differences between the two are approximately $10\%-15\%$ for temperatures below 30~MeV, escalating as the temperature approaches $T=62$~MeV, reaching around a $50\%$ difference. This discrepancy is evident in Fig.~\ref{fig:rate_comparison}, where the most significant deviation occurs for masses around the eV range, aligning with the decoupling temperatures approximately at that value for $\Tend \lesssim 100$~MeV. From this figure, we can also stress the importance of the smooth rate in constraining small axion masses and higher $\Tend$ temperatures.

\subsubsection*{Changes in the exclusion limit of the axion-gluon coupling by considering more LSS data}
We present a comparison of our findings with those in the literature, specifically Refs~\cite{Grin:2007yg, Carenza:2021ebx}. They investigated the parameter space for thermal axions within the framework of KSVZ axions, focusing on the gluon coupling, within the context of LTR cosmology. Both works utilized the axion-pion interaction rate outlined in Ref.~\cite{Hannestad:2005df}, as presented in Eq.~\eqref{eq:rate}, noting its applicability up to temperatures of 62~MeV.

To set their constrained parameter space, Ref.~\cite{Grin:2007yg} used the saturation of the relic abundance from cold dark matter, the ISW effect on the CMB and data from WMAP first year and SDSS measurements of the galaxy power spectrum found in Ref.~\cite{Hannestad:2005df}.  On the other hand, Ref.~\cite{Carenza:2021ebx} updated the parameter space for thermal axions in LTR cosmologies by using CMB observations from Planck 2018 legacy data release~\cite{Planck:2018vyg}, together with BAO measurements from BOSS Data Release 12~\cite{BOSS:2016wmc}, 6dFGS~\cite{Beutler:2011hx} and SDSS-MGS~\cite{Ross:2014qpa}.

\begin{figure}[t!]
    \centering
    \includegraphics[scale=0.48]{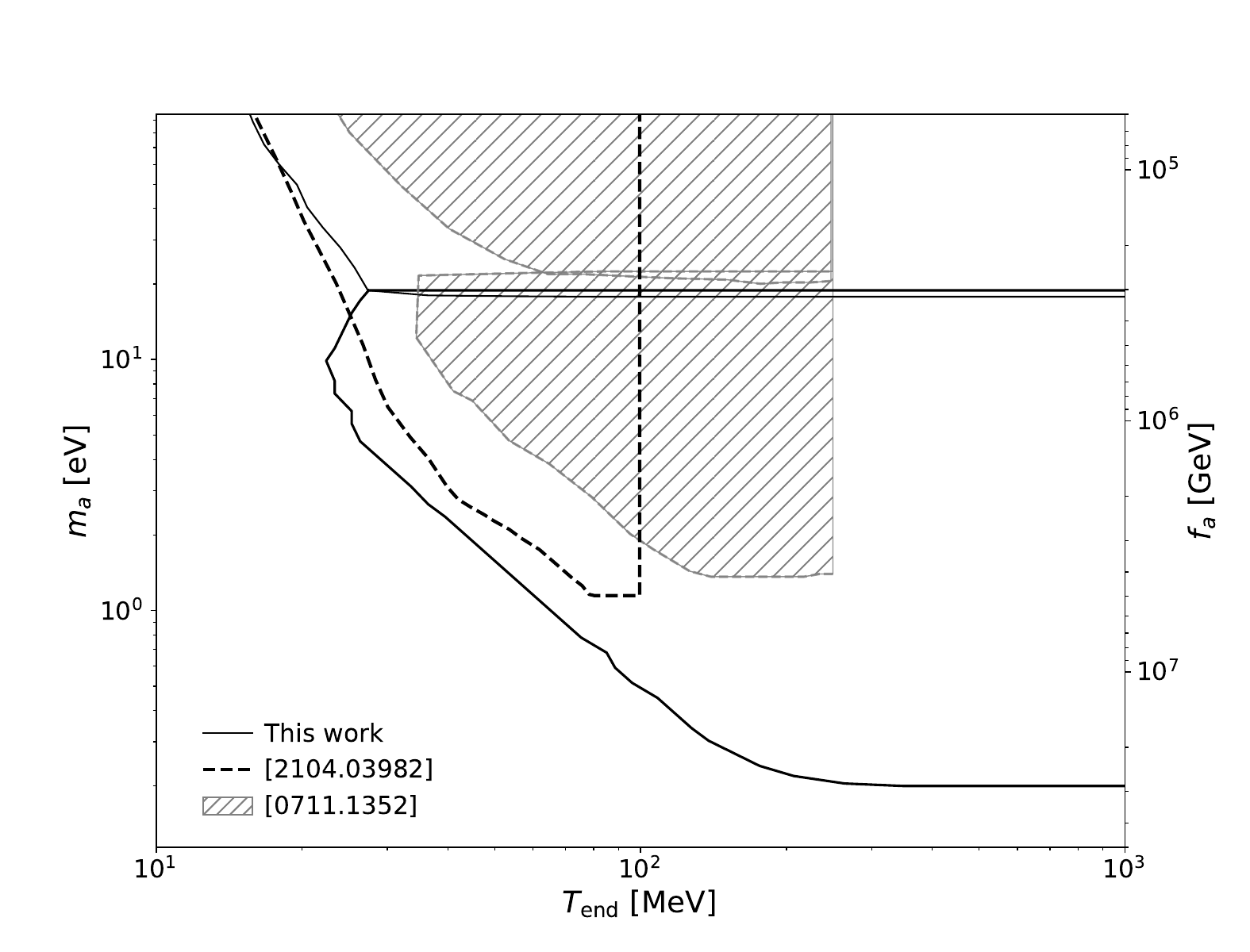}
    \caption{Parameter space as a function of the axion mass/decay constant and the reheating NSC end temperature, $\Tend$, for LTR cosmologies for the axion-gluon coupling. The hatched area corresponds to the limits imposed by Ref.~\cite{Grin:2007yg}. The dashed black line corresponds to the exclusion set by Ref.~\cite{Carenza:2021ebx}, where only masses whose decoupling temperature is less than 62~MeV are shown. Finally, the continuous black line is the bound found in this work, using the smooth rate found in Ref.~\cite{DEramo:2021psx} and the data from Ref.~\cite{Xu:2021rwg}.}
\label{fig:data_comparison}
\end{figure}
In the left panel of Fig.~\ref{fig:data_comparison}, we present a comparative analysis. The excluded parameter space from Ref.~\cite{Grin:2007yg} is depicted in the hatched area, while the limit from \cite{Carenza:2021ebx} is illustrated by the dashed black line, considering only the segment where the rate's validity is acknowledged. Our excluded region, previously presented in Fig.~\ref{fig:bounds_ma_Tend}, is represented by the solid black line. Thus, it becomes apparent that, on the one hand, the smooth rate allows the low-axion-mass region to be confined, where the decoupling temperature is higher than $62$~MeV, and, on the other hand, the incorporation of more robust LSS data allows for a better constraint of the parameter space overall. In particular, the bound on $\Delta N_{\rm eff}$ is refined for sub-eV masses, and in the eV region the LSS data impose significant restrictions, as expected from Fig.~\ref{fig:bounds_22}.

\bibliographystyle{JHEP}

\providecommand{\href}[2]{#2}\begingroup\raggedright\endgroup

\end{document}